\documentclass[twocolumn]{aastex61}
\usepackage{ulem,xcolor}

\newcommand{\Msun}{\mbox{\,$M_{\odot}$}}

\usepackage{epsf}
\usepackage{graphicx}
\usepackage{natbib}
\usepackage{url}
\usepackage{mathrsfs}
\usepackage{amsmath}
\font\smcap=cmcsc10

\newcommand{\kms}{\,km~s$^{-1}$}
\newcommand{\degree}{$^\circ$}
\newcommand{\nai}{Na\,{\smcap i}}
\newcommand{\caii}{Ca\,{\smcap ii}}

\newcommand{\feh}{$\rm[Fe/H]$}
\newcommand{\afe}{$\rm[\alpha/Fe]$}
\newcommand{\fehp}{$\rm[Fe/H]_{phot}$}

\newcommand{\vhel}{$v_{\rm hel}$}

\newcommand{\rproj}{$R_{\rm proj}$}

\accepted{August 2, 2019}
\submitjournal{ApJ}

\shorttitle{First Alpha and Iron Abundance Measurements in M31's Giant Stellar Stream}
\shortauthors{Gilbert et~al.}

\begin{document}
\bibliographystyle{aasjournal}

\title{Elemental Abundances in M31: First Alpha and Iron Abundance Measurements in M31's Giant Stellar Stream\footnote{The data presented herein were obtained at the W.M. Keck Observatory,
which is operated as a scientific partnership among the California
 Institute of Technology, the University of California and the National
Aeronautics and Space Administration. The Observatory was made
possible by the generous financial support of the W.M. Keck
Foundation.}}

\correspondingauthor{Karoline M. Gilbert}
\email{kgilbert@stsci.edu}

\author[0000-0003-0394-8377]{Karoline M. Gilbert}
\affiliation{Space Telescope Science Institute, 3700 San Martin Dr., Baltimore, MD 21218, USA} 
\affiliation{Department of Physics \& Astronomy, Bloomberg Center for Physics and Astronomy, Johns Hopkins University, 3400 N. Charles Street, Baltimore, MD 21218, USA}

\author[0000-0001-6196-5162]{Evan N. Kirby}
\affiliation{California Institute of Technology, 1200 E. California Boulevard, MC 249-17, Pasadena, CA 91125, USA}

\author[0000-0002-9933-9551]{Ivanna Escala}
\affiliation{California Institute of Technology, 1200 E. California Boulevard, MC 249-17, Pasadena, CA 91125, USA}
\affiliation{Department of Astrophysical Sciences, Princeton University, 4 Ivy Lane, Princeton, NJ, 08544, USA}

\author{Jennifer Wojno}
\affiliation{Department of Physics \& Astronomy, Bloomberg Center for Physics and Astronomy, Johns Hopkins University, 3400 N. Charles Street, Baltimore, MD 21218, USA}

\author{Jason S. Kalirai}
\affiliation{Johns Hopkins University Applied Physics Laboratory, MS 200-E181, 11100 Johns Hopkins Road, Laurel, MD 20723, USA}

\author{Puragra Guhathakurta}
\affiliation{UCO/Lick Observatory, Department of Astronomy \& Astrophysics, University of California Santa Cruz, 
 1156 High Street, 
 Santa Cruz, California 95064, USA}

\setcounter{footnote}{11}

\begin{abstract}

We present the first measurements of \feh\ and \afe\ abundances, obtained using spectral synthesis modeling, for red giant branch stars in M31's giant stellar stream. The spectroscopic observations, obtained at a projected distance of 17~kpc from M31's center, yielded 61 stars with \feh\ measurements, including 21 stars with \afe\ measurements, from 112 targets identified as M31 stars.  The \feh\ measurements confirm the expectation from photometric 
metallicity estimates that stars in this region of M31's halo are relatively metal-rich compared to stars in the MW's inner halo: more than half the stars in the field, including those not associated with kinematically identified substructure, have \feh\ abundances $> -1.0$.  
The stars in this field are $\alpha$-enhanced at lower metallicities, while \afe\ decreases with increasing \feh\ above metallicities of \feh\,$\gtrsim-0.9$.  Three kinematical components have been previously identified in this field: the giant stellar stream, a second kinematically cold feature of unknown origin, and M31's kinematically hot halo.  We 
compare probabilistic \feh\ and \afe\ distribution functions for each of the components.  The giant stellar stream and the second kinematically cold feature have very similar abundance distributions, while the halo component is more metal-poor.  Although the current sample sizes are small, a comparison of the abundances of stars in the GSS field with abundances of M31 halo and dSph stars from the literature indicate that the progenitor of the stream was likely more massive, and experienced a higher efficiency of star formation, than M31's existing dSphs or the dEs NGC~147 and NGC~185. 
\end{abstract}


\setcounter{footnote}{0}

\section{Introduction}\label{sec:intro}

Stellar halos provide a record of the earliest stages of a galaxy's formation as well as the mass growth of later epochs. All stages of accretion are represented in the halo: (1) fully phase-mixed stars accreted at early times, (2) stars in distinct tidal streams, and (3) stars in satellite galaxies that will eventually be tidally incorporated into the halo. In addition, the innermost regions of stellar halos preserve a record of the stars formed within the progenitor host halo at very early times, such as stars formed in a proto-disk and later heated into the halo \citep[e.g.,][]{purcell2010}.

The stellar halos of the Milky Way (MW) and Andromeda (M31) galaxies thus provide observational probes of the formation and accretion histories of $L_*$ galaxies.  
Wide-field imaging has revealed a wealth of structure in the stellar halos of both galaxies, including massive tidal debris features such as the Sagittarius Stream \citep[e.g.,][]{ibata1994,mateo1998sgr,yanny2000,majewski2003} in the MW and the giant stellar stream (GSS) in M31 \citep{ibata2001a}. 
Significant observational and theoretical effort has been expended to determine the properties of the progenitors of the brightest tidal debris structures in the MW and M31 halos and model the interactions that produced them, and to constrain the relative contribution of accreted stars and in-situ star formation to the stellar halos of MW and M31 mass galaxies.

Stellar chemical abundances can provide powerful observational constraints, as the combination of \feh\ and \afe\ ratios can be used to infer the star formation history of a population \citep[e.g.,][]{wheeler1989,gilmore1991,gilmore98}. For example, \citet{shetrone2001} and \citet{venn2004} showed that the 
classical MW satellites have more steeply declining tracks of \afe\ as a function of \feh\ than MW halo stars. 
This difference in abundance pattern can be attributed to a more gradual early star formation history for the satellites. In the satellite galaxies, massive supernovae, which produce abundant $\alpha$-elements (O, Ne, Mg, Si, S, Ar, Ca and Ti), enriched the stellar population only to \feh\,$\sim -2.3$ before Type Ia supernovae, whose ejecta are Fe-rich, began to depress the \afe\ ratio. On the other hand, the progenitors of the MW halo had such vigorous early star formation that they reached \feh\,$\sim -1$ before Type Ia supernovae began to explode. 

Thus, in addition to providing information on the star formation history and chemical evolution of surviving satellites and the progenitors of intact tidal debris features, \feh\ and \afe\ abundances provide a means of constraining the properties of the progenitors of the underlying phase-mixed stellar halo.
Simulations by \citet{bullock2005} and \citet{johnston2008} showed that the accretion history can imprint a strong chemical signature on a MW- or M31-like halo. Stars from the most massive disrupted satellites are expected to dominate the inner halo and comprise the majority of the halo's stellar mass. These massive accretion events typically occurred over 9 Gyr ago. As a result, the chemical abundances of the majority of halo stars are expected to reflect the star formation histories of rapidly formed galaxies that were shut off early. Such a stellar population is $\alpha$-enhanced (\afe\,$\sim +0.3$) and not particularly metal-poor ($\langle$\,[Fe/H]$\rangle \sim -1$ to $-0.5$). Less massive satellites were accreted later and deposited debris further from the center of the host potential \citep[see also][]{cooper2010, tissera2013}.

Recent improvements in the observational constraints on the mass distribution and time of infall of accreted satellite galaxies in the Milky Way have come from datasets that cannot (currently) be replicated for M31, including {\it Gaia}, APOGEE, and ratios of blue horizontal branch stars and blue stragglers \citep[e.g.,][]{deason2015, belokurov2018, helmi2018, haywood2018, dimatteo2018, mackereth2019, fattahi2019, myeong2019}. However, the analysis of recent stellar halo simulations has led to several suggested methods and diagnostics for constraining these quantities using abundance measurements (\feh\ and \afe) of halo stars \citep[e.g.,][]{lee2015, deason2016, dsouza2018MNRAS, monachesi2019}.  
Measurements of \afe\ and \feh\ ratios thus provide a key observational dimension for 
deciphering the likely origins of the stellar populations in M31's halo.

\subsection{Chemical Abundances in M31}

Studies of M31's stellar halo have made tremendous progress, 
from the discovery of M31's extended stellar halo \citep{irwin2005,guhathakurta2005,gilbert2006} to characterization of its global properties  \citep{kalirai2006halo, ibata2007, koch2008, mcconnachie2009, gilbert2012, gilbert2014, ibata2014}. Spectroscopy of stars in M31's halo has determined the fraction of stars in tidal streams in individual fields and characterized the kinematical properties of tidal debris features and the halo as a whole \citep[e.g.,][]{ibata2004, gilbert2007, chapman2008, gilbert2009gss, gilbert2009a, gilbert2012, dorman2012, gilbert2018}. 

However, little is known about the chemical abundances of stars in M31's halo beyond estimates of \feh\ based on comparisons of photometric measurements of stars to theoretical isochrones or measurements of calcium absorption lines \citep{kalirai2006halo, chapman2006, koch2008, kalirai2009, richardson2009, tanaka2010, ibata2014, gilbert2014, ho2015}. This stands in stark contrast to the detailed chemical abundance measurements made for stars in the MW's halo and satellites \citep{cayrel2004,barklem2005,cohen2008,lai2008,roederer2009,frebel2010,shetrone2001,venn2004,kirby2009,kirby2010,cohen2010}. 
Obtaining measurements of \feh\ and \afe\ for stars in M31's most prominent tidal debris feature, the GSS\@, has taken on greater significance given recent suggestions that it may have been been formed by a major merger \citep{hammer2018, dsouza2018} instead of a minor merger as previously assumed \citep[e.g.,][]{font2006b, fardal2006}.  If true, this has significant implications for our understanding of the evolution of the M31 system, affecting interpretations of observations throughout M31's disk as well as halo.

At the distance of M31, the top several magnitudes of the RGB can be observed with modest aperture ($\sim 4$-m class) ground-based telescopes, enabling photometric metallicity estimates to be obtained throughout the stellar halo \citep{ibata2014}. Photometrically derived metallicities agree with spectroscopic metallicities for old, single-age populations, but not for populations with more than 15\% intermediate-age stars \citep{lianou2011}. Deep imaging with HST has demonstrated that stars in multiple fields in M31's inner halo (to \rproj\,$= 35$~kpc) span a range of ages, as large as 5\,--\,12~Gyr \citep{brown2006,brown2008}. 
Several dwarf spheroidal satellites in M31 
also show a significant range of stellar ages \citep{weisz2014a,weisz2014}.

Metallicity can also be estimated from the equivalent widths (EWs) of very strong metal absorption lines, like the \caii\ triplet (CaT) at $\sim8500$~\AA\@.   
The EW-metallicity relation 
is empirically calibrated
against abundance standards
\citep[e.g.,][]{armandroff1991,battaglia2008,dacosta2016}.  The CaT is a strong spectral feature 
identifiable even in low-SNR spectra, leading to its wide adoption for
measuring [Fe/H] \citep[e.g.,][]{tolstoy2001,helmi2006,battaglia2008,deboer2012}.  
However, the precision of the CaT metallicity estimates are limited by the precision of the EW measurement of only two or three lines.  Moreover, at the line-of-sight velocities typical for stars in M31's stellar halo, these lines are frequently affected by night sky lines.  This, in combination with the relatively low SNR achievable for spectra of red giant stars at M31's distance, results in large uncertainties in CaT metallicity estimates \citep[e.g.,][]{gilbert2014}.

An alternative use of medium-resolution spectra is full spectral synthesis, which can yield a measurement of \feh\ from a large number of Fe lines directly. Crucially, it also allows a measurement of \afe, which cannot be constrained or derived from the above methods. Spectral synthesis was recently applied to M31 red giants for the first time, yielding measurements of \afe\ and \feh\ for 226 red giants in nine M31 satellite galaxies as well as four halo stars identified in three M31 dSph fields \citep{vargas2014, vargas2014apjl}.

We have undertaken a program to measure \feh\ and \afe\ of M31 stars associated with all stages of halo formation: the ``smooth'' halo, accreted or formed within M31's potential long ago; stellar streams, accreted recently; and M31's surviving satellites, which have yet to be accreted into the halo. The first work in this series, \citet{escala2019}, presents the application of the spectral synthesis method to $R\sim2500$~spectra and measures the \feh\ and \afe\ abundances for 11 stars in a relatively smooth M31 halo field at \rproj\,$=23$~kpc.  In this contribution, we present the first \feh\ and \afe\ abundances of a tidal debris feature in M31's halo. The spectroscopic field analyzed here is located in the inner regions of M31's GSS. 

This paper is organized as follows. Section~\ref{sec:data} provides a brief overview of the spectroscopic dataset and the known properties of the GSS field. Section~\ref{sec:abund} summarizes the spectral synthesis method and presents the chemical abundance measurements. Section~\ref{sec:mdfs} describes the derivation of probabilistic abundance distributions for each M31 component present in the field. Section~\ref{sec:context} presents the results in the context of M31's stellar halo and dwarf satellite population.  Section~\ref{sec:disc} discusses the implications of the measured abundance distributions for different merger scenarios
and revisits the possible origins of the second kinematically cold component in the field.  Section~\ref{sec:conc} summarizes the results.

\section{Data}\label{sec:data}

\subsection{Spectroscopic Data}\label{sec:specdata}

\begin{figure*}[tb]
\includegraphics[width=\textwidth]{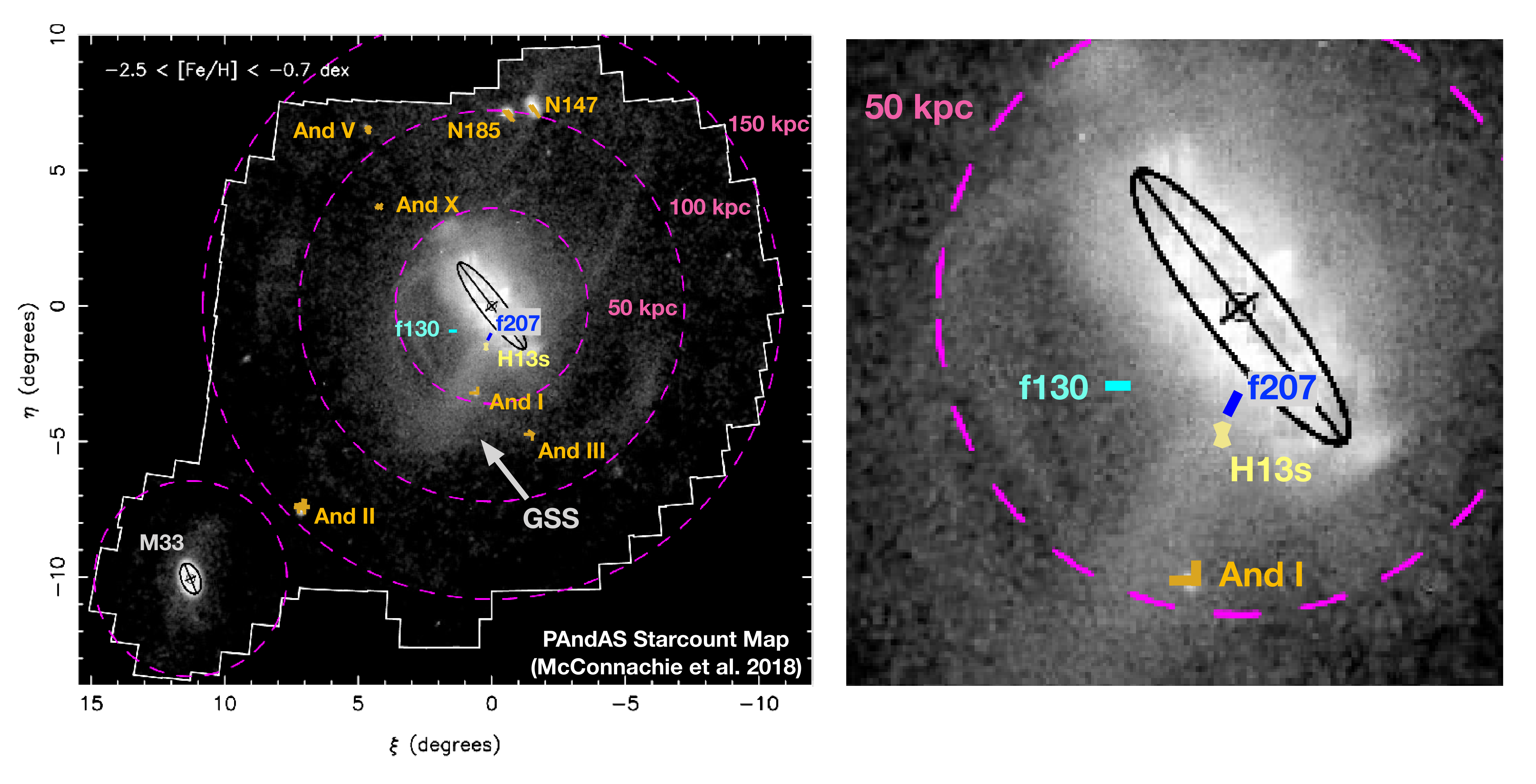}
\caption{
(\textit{Left:}) Location of the GSS spectroscopic mask analyzed here (`f207\_1a', blue point and label) in the full context of M31's stellar halo as seen by the PAndAS survey \citep{mcconnachie2018}, the extent of which is shown by the white outline.  The underlying star count map \citep[Figure~11 of][]{mcconnachie2018} shows the surface density of stars within a range of $-2.5 < $\fehp\,$<-0.7$, where \fehp\ was derived from a comparison of the star's location in a color-magnitude diagram with theoretical isochrones.  Other relevant spectroscopic fields are also shown, including fields with existing \feh\ and \afe\ abundance measurements to which the GSS measurements are compared (orange points and labels). (\textit{Right:}) A close-up view of the location of relevant spectroscopic masks in the inner regions of M31's halo. 
}
\label{fig:roadmap}
\end{figure*}

The field analyzed here is located on M31's GSS at a mean projected distance from M31's center of 17~kpc (Figure~\ref{fig:roadmap}).  We observed this field with the DEIMOS Spectrograph \citep{faber2003} on the Keck~II 10-m telescope on 2017 Oct 22, for a total integration time of 6.4 hours in 0.7\arcsec--0.9\arcsec\ seeing.  

The spectroscopic mask for these deep observations, `f207\_1a,' was designed using results from a mask previously observed for one hour 
as part of the SPLASH survey \cite[field `f207,' Figure~1 of][]{gilbert2009gss}.  Stars previously identified as likely to be red giant branch stars in M31 were included on the mask with highest priority, while stars previously identified as likely Milky Way foreground stars or background galaxies were omitted.  

Targets were chosen using a catalog \citep{kalirai2006gss} derived from $g'$ and $i'$ band 
imaging obtained with the MegaCam instrument on the 3.6-m Canada-France-Hawaii 
Telescope (CFHT)\footnote{MegaPrime/MegaCam
is a joint project of CFHT and CEA/DAPNIA,
at the Canada-France-Hawaii Telescope
which is operated by the National Research Council of Canada, the Institut
National des Science de l'Univers of the Centre National de la Recherche
Scientifique of France, and the University of Hawaii.}.  The calibrated photometry was transformed to Johnson-Cousins $V$ and $I$ band magnitudes using observations of \citet{Landolt1992} photometric standard stars.

The spectroscopic mask was observed with 
the 1200~line~mm$^{-1}$ grating, a slit width of 0.7\arcsec, 
and a central wavelength setting of 7800~\AA\@.  
This results in a dispersion of 0.33~\AA~pix$^{-1}$, a resolution of 1.2~\AA\ FWHM,
and a wavelength range of 
$\lambda\lambda\sim$~6450\,--\,9150~\AA.   

The spectra were reduced using the {\tt spec2d} (flat-fielding, night-sky emission line removal, and extraction
of one-dimensional spectra) and {\tt spec1d} (redshift measurement) software developed
at the University of California, Berkeley \citep{cooper2012,newman2013} with {\tt spec2d} modifications by \citet{Kirby15a}.  The line of sight velocities were measured by cross-correlating the observed spectra with a suite of stellar templates provided by \citet{simon2007}.
The measured line of sight velocities were transformed to the heliocentric frame and
were corrected for imperfect centering of the star in the slit, which is measured by cross-correlating each stellar spectrum with a telluric template, using regions of strong telluric absorption \citep{sohn2007,simon2007}. 

\begin{figure}[tb]
\includegraphics[width=0.5\textwidth]{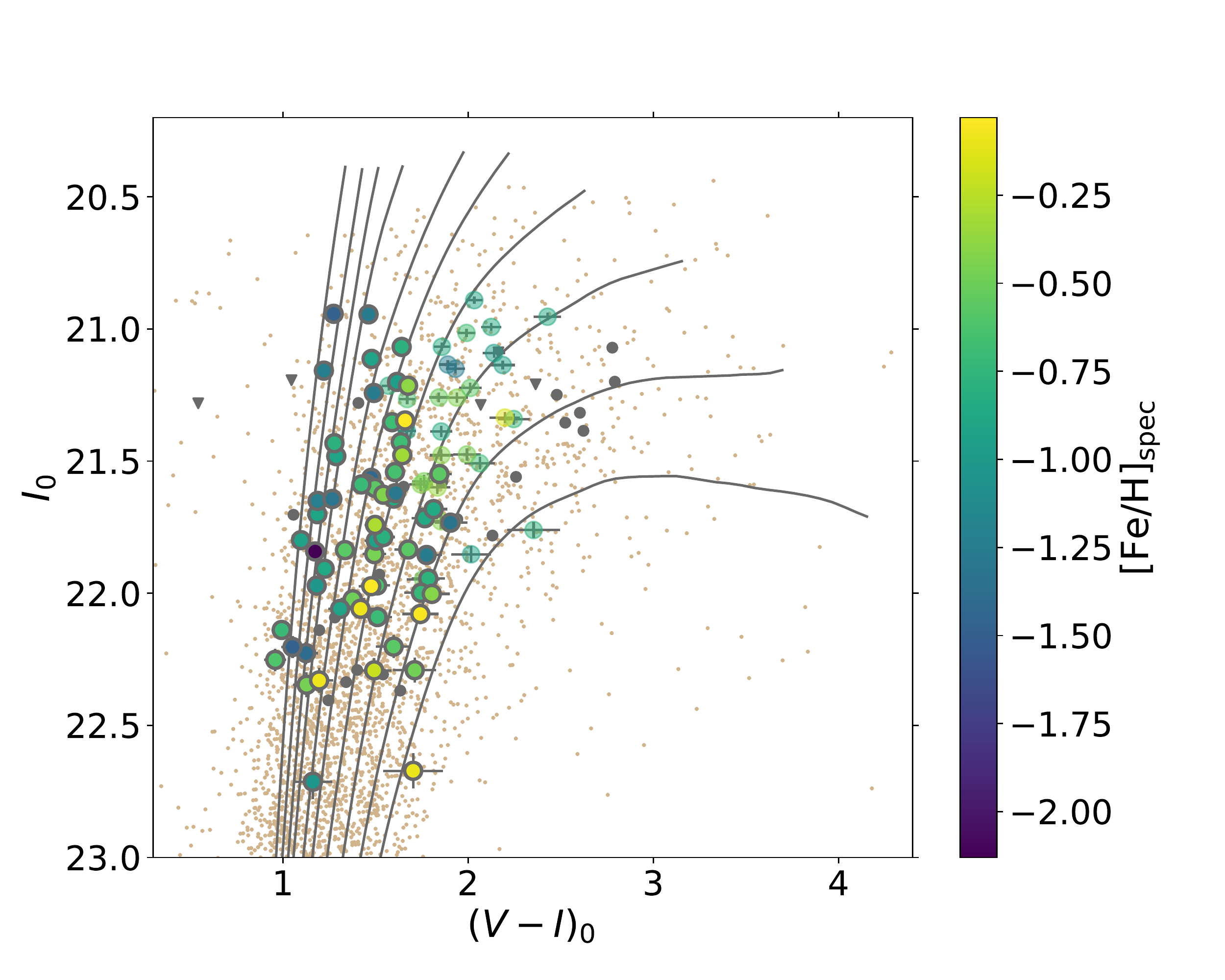}
\caption{
Color-magnitude diagram showing the location of spectroscopic targets with measured line-of-sight velocities. Stars at least three times more likely to be red giants in M31 are denoted with circles, while stars that do not meet this criterion are denoted with triangles  (Section~\ref{sec:specdata}).  M31 red giant branch stars with \feh\ measurements derived from spectral synthesis are colored according to the derived \feh.  Stars without \feh\ measurements are shown in gray.  Stars with evidence of TiO features in their spectrum (transparent points lacking outlines) are not considered in the analysis, as the reliability of the spectral synthesis abundance measurements has not been established.   Small tan points show the distribution of all stellar objects in the photometric catalog from which the spectroscopic slitmask was designed.  Twelve Gyr, \afe\,=\,0.0 isochrones are shown for reference \citep[\feh\,=\,$-2.31$, $-1.84$, $-1.53$, $-1.31$, $-1.01$, $-0.83$, $-0.61$, $-0.40$, $-0.20$, 0.0;][]{vandenberg2006}. 
}
\label{fig:cmd}
\end{figure}

The final M31 dataset presented here contains only stars that are more than three times as likely to be M31 RGB stars than foreground MW stars.  This likelihood is computed using a set of four empirical photometric and spectroscopic diagnostics established by \citet{gilbert2006}: (1) line-of-sight velocity, 
(2) strength of the \nai\ doublet absorption line as a function of
($V-I$) color, 
(3) location in the ($I$, $V-I$) color-magnitude diagram, 
and (4) comparison of CaT-based and photometric metallicity estimates.  Of the 117 stellar spectra with measured velocities, 112 are securely classified as M31 RGB stars (Figure~\ref{fig:cmd}). 
The reader interested in 
further details is referred to  
\citet{gilbert2012} for a concise description of the classification of M31 and MW stars.

\subsection{Properties of the Field}\label{sec:field}

\begin{figure}[tb]
\includegraphics[width=0.5\textwidth]{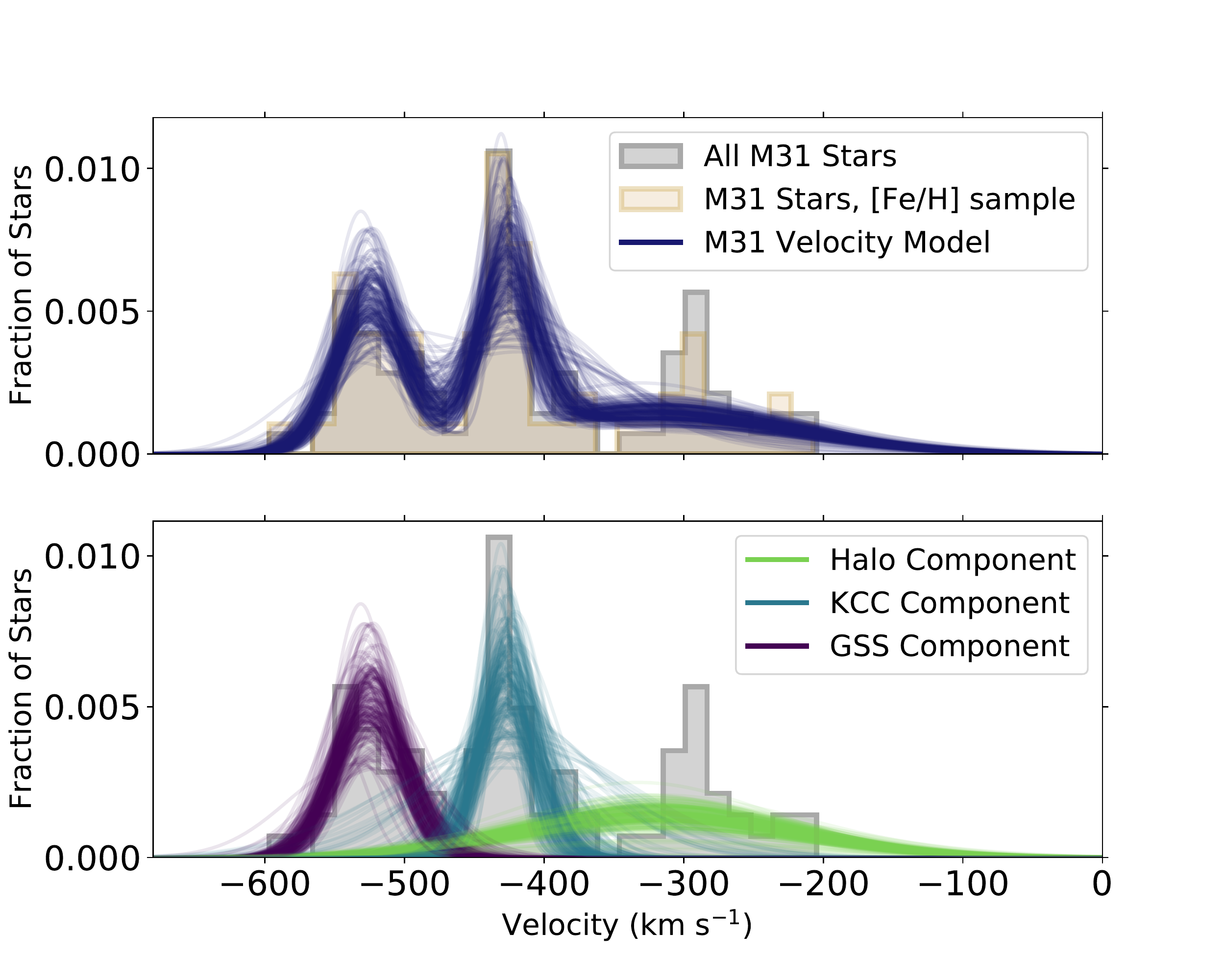}
\caption{
Velocity distributions of all M31 stars in the field with successful velocity measurements, as well as with \feh\ measurements that pass all quality cuts (Section~\ref{sec:abund}). Overlaid are 200 samples of the parameterized velocity distribution model for this field, drawn from the MCMC chain \citep{gilbert2018}.  The top panel shows the full M31 velocity model for the field, which includes two kinematically cold components [the GSS and a component of unknown origin (the KCC)]
as well as a kinematically hot halo (Section~\ref{sec:field}).  
The bottom panel shows the model velocity distributions for the individual components.  The mean velocity of the GSS is \vhel\,$= -525$\kms, the mean velocity of the KCC of unknown origin is 
\vhel\,$= -427$\kms, and the systemic velocity of M31 is $v_{\rm sys} = -300$~\kms.  The GSS, KCC, and halo components each comprise one-third of the M31 RGB population in this field (Appendix~\ref{sec:app_velmodel}).
}
\label{fig:vel_model}
\end{figure}

The field presented here is known to be comprised of multiple M31 components (Figure~\ref{fig:vel_model}): 
(1) the GSS, which has been identified and characterized both kinematically and 
spatially, (2) a kinematically cold (velocity dispersion $\sim 20$~\kms) 
component (KCC) of unknown origin, 
(3) a kinematically hot halo represented by stars with a large velocity dispersion.   

The mean velocity and velocity dispersion of
stars in the GSS and KCC in this field were initially measured by \citet{gilbert2009gss}.  More recently,
\citet{gilbert2018} fit a Gaussian Mixture Model to the entire SPLASH M31 halo dataset,
using MCMC techniques to obtain posterior probability distribution functions for
the mean velocity and velocity dispersion of M31's halo as a function of projected distance from M31, as well as for all
M31 kinematical components previously identified in each of the spectroscopic fields (Appendix~\ref{sec:app_velmodel}).  The mean velocity, velocity dispersion, and fraction of stars in each component are reported in Table~\ref{components_table}.  
Figure~\ref{fig:vel_model} displays a representation of the M31 velocity model for the field presented here, based on fitting the velocity distribution of stars in the original f207 spectroscopic mask. Multiple draws from the MCMC chains are shown to illustrate the underlying uncertainties in the model parameters.  The halo component of the model was constrained using a significantly larger sample of stars in M31's stellar halo, drawn from additional spectroscopic fields at comparable distances from M31's center.  

The velocity distribution of stars in our \feh\ sample is consistent with the velocity model for this field computed by \citet{gilbert2018}, as is expected given the large overlap in targets between the original and deep spectroscopic masks. While the two kinematically cold features in this field each have relatively small velocity dispersions, there is still expected to be some overlap in the velocities of stars belonging to different components, in particular between stars belonging to the underlying kinematically hot halo and stars belonging to the KCC of unknown origin.

\begin{figure}[tb]
\includegraphics[width=0.5\textwidth]{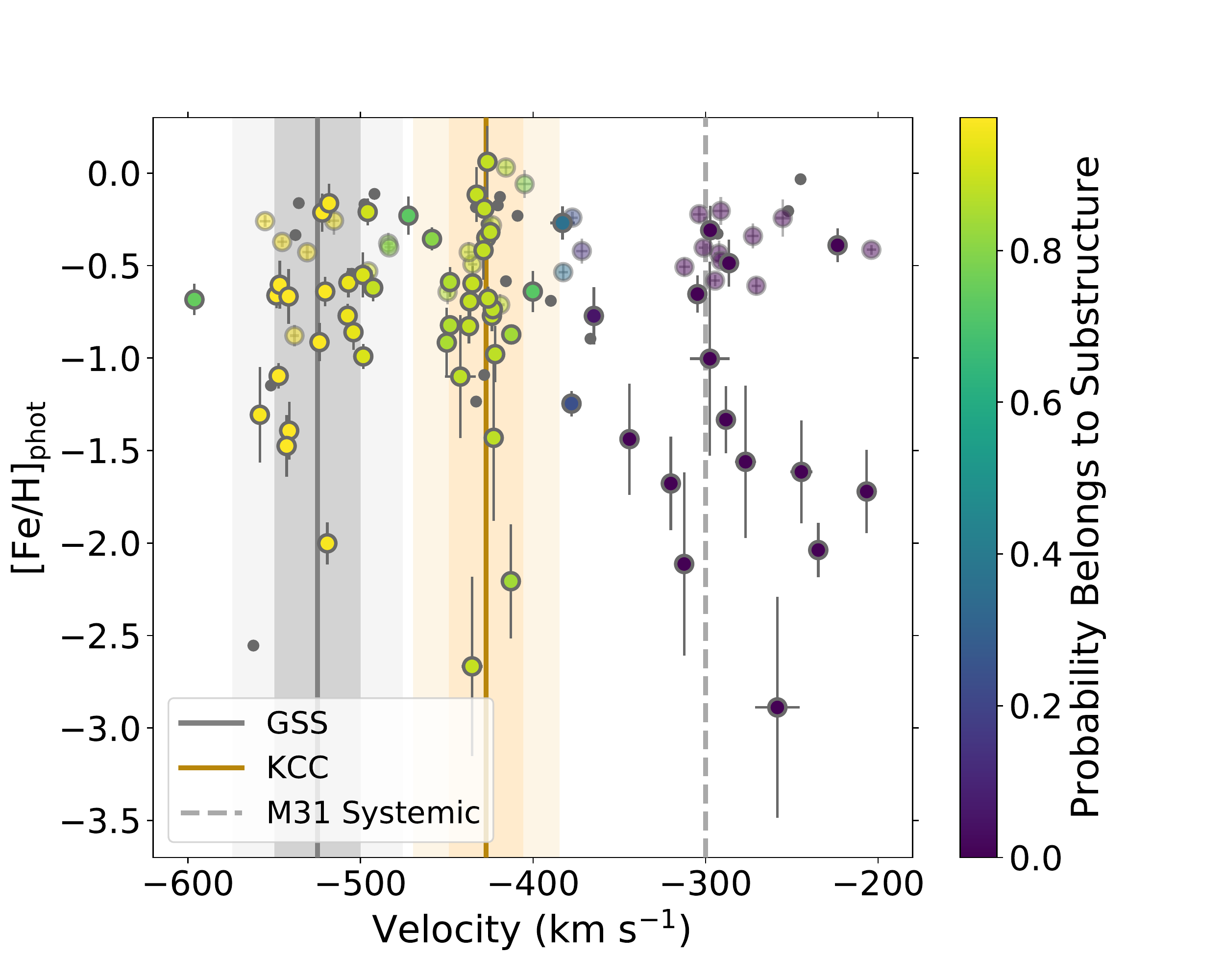}
\caption{
Photometric 
\fehp\ estimates (the basis of all previous estimates of the metallicity of stars in M31's GSS) as a function of heliocentric velocity 
for all M31 stars in the field. 
Stars with successful \feh\ measurements derived from the spectral synthesis fitting (Section~\ref{sec:abund}) are color-coded by the 
probability that the star belongs to either of the two tidal debris features (Section~\ref{sec:field}) identified in the field ($p_{\rm subst} = 1 - p_{\rm halo}$).  
Stars with evidence of TiO features in their spectrum, which are not included in the analysis, are denoted by transparent points, 
and stars without successful \feh\ measurements are denoted by small grey points.  
The solid lines show the 50th percentile mean velocities ($\mu$; see Figure~\ref{fig:corner}) of the 
two kinematically cold components in this field (grey for the GSS and beige for the KCC); darker and lighter shadings denote $\pm 1$ and $\pm 2$ times the 
50th percentile velocity dispersion ($\sigma$; see Figure~\ref{fig:corner}) of each component, respectively.  
The systemic velocity of M31 
is denoted by the grey dashed line.  Stars likely associated with the GSS and KCC appear to span similar ranges of \fehp, while the majority of metal-poor stars appear likely to belong to the underlying inner halo component.
}
\label{fig:fehphot_vs_vel}
\end{figure}

The two kinematical components observed in this field (`f207') are also observed in an adjacent field on the GSS \citep[`H13s,' Figure~\ref{fig:cmd};][]{kalirai2006gss}. It is not known whether the two 
kinematically cold features (the GSS and the KCC) arise from the same accretion event.  Previous
observations indicated a potential link in both chemistry and kinematics between the two features.  The mean velocities of the two components
follow a similar trend as a function of projected distance from M31 \citep[Figure~16 of][]{gilbert2009gss}, 
and the photometrically-derived \fehp\ distributions of the two features are quite similar \citep[Figure~17 of][]{gilbert2009gss}.  

There appears to be a small excess of stars at $\sim -300$~\kms.  This excess was noted in the velocity distribution of stars observed on the original mask, but it was found to not be statistically significant \citep{gilbert2009gss}.  A similar feature has not been observed in any adjacent spectroscopic fields \citep{gilbert2009gss}.  

Figure~\ref{fig:fehphot_vs_vel} shows the photometric metallicity estimates for all M31 stars in the field as a function of line of sight velocity.  Comparisons of the magnitudes and colors of stars with theoretical isochrones form the basis of all previous estimates of the
metallicity of stars in M31's GSS \citep[]{ibata2001a, mcconnachie2003, brown2006, gilbert2009gss, gilbert2014, ibata2014, conn2016}.     
Photometric metallicities (\fehp) are based on a comparison of the location of the star in a color-magnitude diagram
with 12.6~Gyr, \afe\,$=0.0$~dex theoretical isochrones \citep{vandenberg2006}; error bars reflect the uncertainty in \fehp\ due to the photometric uncertainties.  
The probabilities that each star belongs to substructure were calculated using the 50th percentile values for the model parameters in this field \citep[Appendix~\ref{sec:app_velmodel};][]{gilbert2018}.       
We analyze these two features separately
below and discuss the implications of our analysis 
in the context of previous observational and theoretical work in Section~\ref{sec:disc}.

\begin{deluxetable*}{rrrrrrr}
\tablecolumns{7}
\tablewidth{0pc}
\tablecaption{Properties of the Three Components in the 17 kpc GSS Field.\label{components_table}}
\tablehead{Component & \multicolumn{3}{c}{Velocity Model\tablenotemark{a}} & \multicolumn{3}{c}{[Fe/H] Distribution\tablenotemark{b}} \\
 & \multicolumn{1}{c}{Mean Velocity} & \multicolumn{1}{c}{Velocity Dispersion} & \multicolumn{1}{c}{Fraction} & \multicolumn{3}{c}{Percentiles} \\
& \multicolumn{1}{c}{(\kms)} & \multicolumn{1}{c}{(\kms)} & & \multicolumn{1}{c}{50th} & \multicolumn{1}{c}{16th} & \multicolumn{1}{c}{84th}
}
\startdata
GSS & $-524.9\pm 4.4$ & $24.5^{+3.9}_{-3.2}$ & $0.33\pm 0.05$ & $-0.75$ & $-1.12$ & $-0.36$ \\
KCC & $-427.3^{+5.4}_{-4.6}$ & $21.0^{+7.4}_{-4.8}$ & $0.32^{+0.07}_{-0.06}$ & $-0.61$ & $-1.00$ & $-0.17$ \\  
Halo & $-319.6^{+4.4}_{-4.2}$ & $98.1^{+5.3}_{-5.0}$ & 0.35 & $-0.94$ & $-1.37$ & $-0.52$ \\
\enddata
\tablenotetext{a}{Quoted values (uncertainties) are the 50th (16th and 84th) percentiles of the marginalized 1-dimensional posterior probability distribution functions from \citet{gilbert2018}, summarized in Section~\ref{sec:field} and Appendix~\ref{sec:app_velmodel}. }
\tablenotetext{b}{Quoted values are the stated percentiles of the cumulative probabilistic \feh\ distribution function, computed using the 50th percentile parameter values for the velocity model as described in Section~\ref{sec:compute_prob_mdfs}.  These values do not account for known systematic biases against measuring \feh\ for metal-rich stars (Section~\ref{sec:mdf_biases}).}  
\end{deluxetable*}

\section{\feh\ and \afe\ Abundance Distributions}\label{sec:abund}

We measure \feh\ and \afe\ for each star by comparing the observed spectrum against a large grid of synthetic spectra to find the model spectrum with the chemical composition that best matches the observed spectrum, following the methodology established by \citet{kirby2008a,kirby2010}.  This technique makes use of lines that are weak and/or blended in moderate- or low-resolution data, leveraging all of the metallicity information in the spectrum simultaneously rather than relying on individual line detections.  This technique enables measurements of abundances from relatively low-SNR spectra.  Because spectral synthesis measures Fe abundances directly from Fe lines and does not rely on intermediary calibrants (in contrast to empirical methods, like CaT-based metallicities), it is applicable over an arbitrary range of metallicity.  Furthermore, the  $\chi^2$ minimization routine that compares the observed spectrum to synthetic spectra can be instructed to operate on just one atomic species. Hence, \afe\ can be measured separately from \feh.  This technique has the potential to provide significantly higher fidelity [Fe/H] estimates, with fewer assumptions, than achievable via CMD-based metallicity estimates or estimates of [Fe/H] derived from measurements of the CaT\@.  Moreover, [$\alpha$/Fe] abundances cannot be reliably determined for RGB stars by other means. 

We employ the synthetic spectral grids of \citet{kirby2008a,kirby2010} and \citet{Kirby2011d}. 
The wavelength-calibrated, sky-subtracted, one-dimensional spectra are shifted to the rest frame, and an initial continuum normalization is performed.  The resulting spectra are
compared to the grid of synthetic spectra to simultaneously fit for
effective temperature (T$_{\mathrm{eff}}$) and [Fe/H]. 
The determined values for these parameters are then held constant as the average [$\alpha$/Fe] (computed from regions of the spectrum sensitive to Mg, Si, Ca, and Ti) is determined. The original spectrum is then divided by the best-fit synthetic spectrum in order to refine the determination of the continuum.  These steps are iterated until the continuum converges.
The revised spectrum, with the final continuum normalization, is then refit to determine [Fe/H].  As in the previous steps, [Fe/H] is then held constant to determine [$\alpha$/Fe].  

\begin{figure}[tb]
\includegraphics[width=0.5\textwidth]{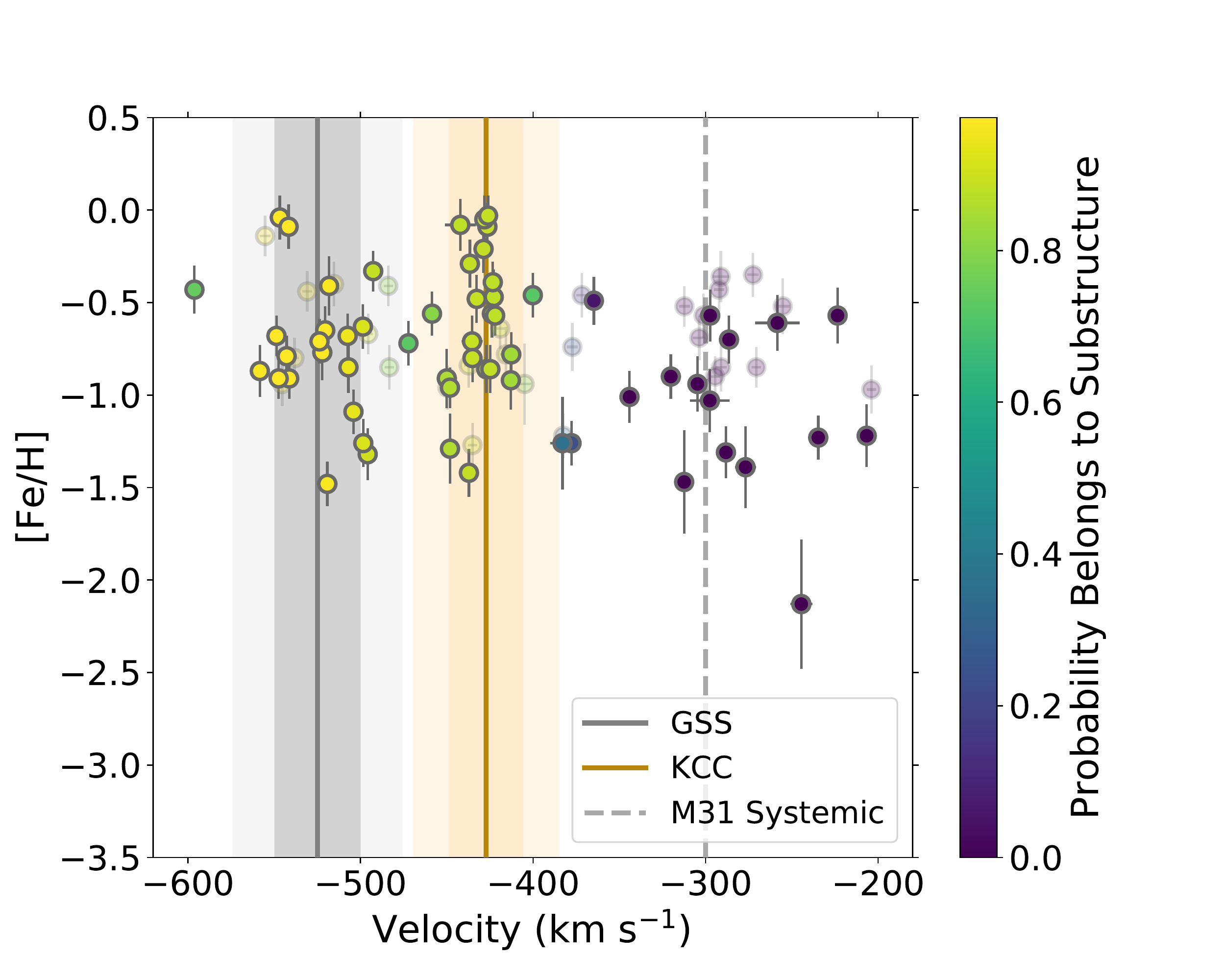}
\includegraphics[width=0.5\textwidth]{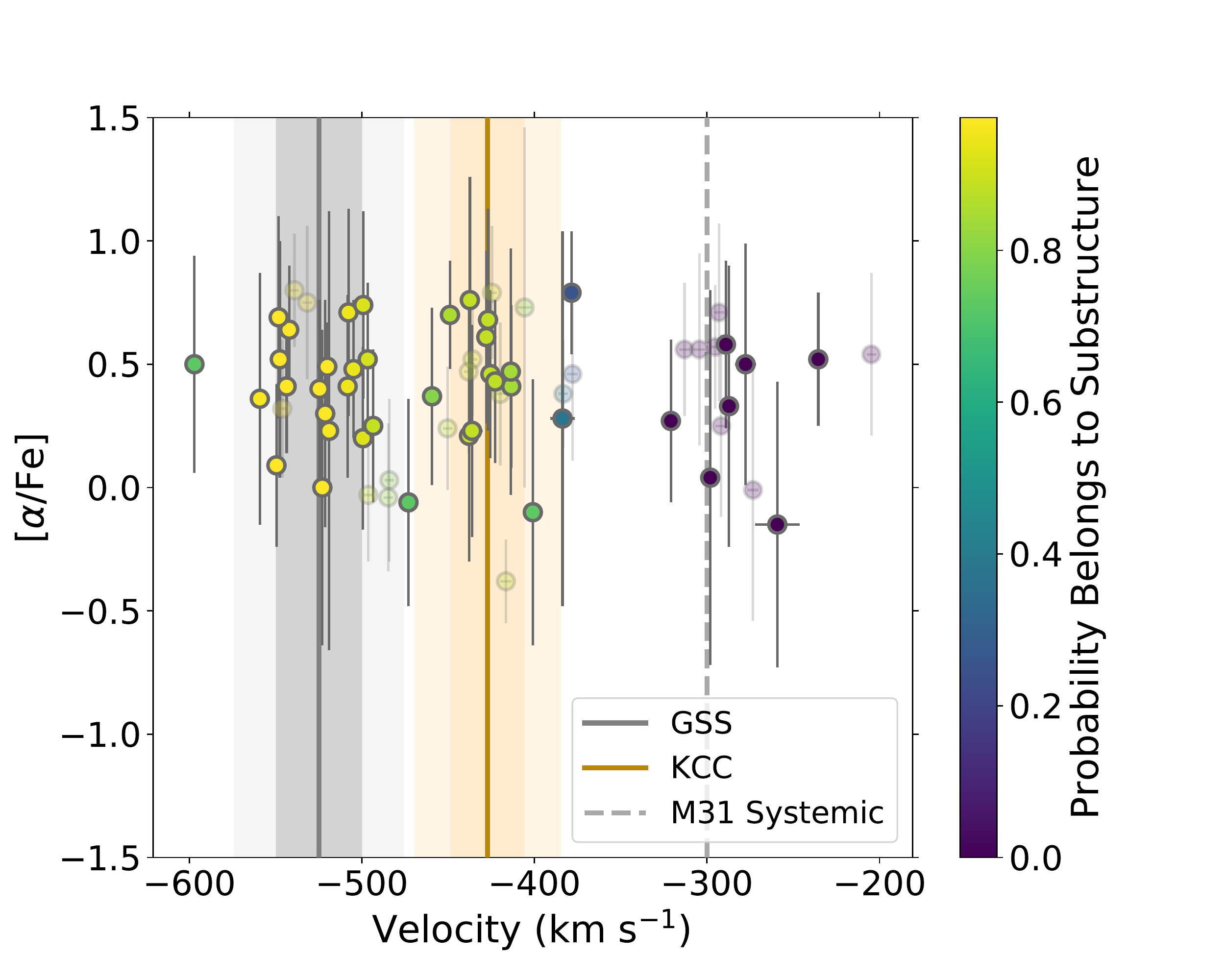}
\caption{Distributions of
\feh\ ({\it top}) and \afe\ ({\it bottom}), derived from spectral synthesis (Section~\ref{sec:abund}), as a function of heliocentric velocity 
for all M31 stars in the field with measured abundances.  As in Figure~\ref{fig:fehphot_vs_vel}, points are color-coded by the 
probability that the star belongs to one of the two tidal debris features identified in the field (Section~\ref{sec:field}).  
Error bars show the combined random and systematic uncertainties for each measurement.
Transparent points denote cool RGB stars with TiO absorption in their spectra. These stars are excluded from the remainder of the analysis (Section~\ref{sec:abund}).  Stars with velocities consistent with the two tidal debris features appear to span a similar range of both \feh\ and \afe\ abundances.  
}
\label{fig:data_feh_vs_vel}
\end{figure}

Random uncertainties in the fit parameters are estimated from the diagonal elements of the covariance matrix. The total uncertainties in the \feh\ and \afe\ estimates are the quadrature sum of the random uncertainty and the systematic error in \feh\ and \afe\ ($\sigma_{\rm sys} = 0.101$ and 0.084, respectively), measured as described by \citet{kirby2010}.  

\citet{kirby2008a, kirby2010} demonstrated that with the spectral synthesis technique, \feh\ and \afe\ of red giant branch stars can be measured to a precision of 0.2 dex from spectra with SNR $\sim15$\,\AA$^{-1}$ obtained with the DEIMOS 1200 line mm$^{-1}$ grating.  
\citet{vargas2014} applied this same technique to DEIMOS spectra of RGB stars in M31 dSphs.  They tested the retrieval of \feh\ and \afe\ using synthetic spectra degraded to the resolution and SNR of the M31 sample.  They found that the recovered parameters remained accurate over the range of \feh\ values covered by our sample, even at SNR $< 15$\,\AA$^{-1}$ for \afe\ and SNR $<8$\,\AA$^{-1}$ for \feh, albeit with decreased precision at low SNR\@.  

Of the 112 M31 RGB stars in the sample, 98 have converged \feh\ measurements from the spectral synthesis analysis, and 77 have converged \afe\ measurements.   We require a well constrained minimum in the $\chi^2$ contours for each of the parameters $T_{\rm eff}$, \feh, and (for \afe\ measurements) \afe. This removes 8 and 14 stars from the \feh\ and \afe\ samples, resulting in 90 and 63 stars with \feh\ and \afe\ measurements, respectively.  
The 22 stars with failed \feh\ measurements or removed from the sample due to insufficiently constrained $\chi^2$ contours are likely to be relatively metal-rich based on their colors and magnitudes: 16 have \fehp\ estimates greater than the median \feh\ of the final sample.

TiO absorption features were not included in the line lists used to produce the model spectra, and these features can be present throughout much of the wavelength range of the observed spectra.  Since we have not validated the fidelity of abundance measurements made with the current library of model spectra for stars with TiO features, we exclude all stars with signatures of TiO absorption in their spectra (at $\lambda\lambda\sim 7050$\,--\,7250\AA\@) from this analysis.  Excluding stars with TiO absorption features results in 61 stars with \feh\ measurements and 41 stars with \afe\ measurements. 

Figure~\ref{fig:data_feh_vs_vel} displays the \feh\ and \afe\ measurements as a function of the line-of-sight velocities.  Stars with line-of-sight velocities consistent with the GSS and the KCC appear to have similar mean \feh\ and \afe\ abundances, and to span a similar range in \feh\ and \afe.  This will be explored further in Section~\ref{sec:mdfs}.  

Figure~\ref{fig:data_feh_vs_vel} also shows the locations of stars with clear TiO features (transparent points).  These stars span the full range of line-of-sight velocities of M31 RGB stars in the field.  There are twice as many TiO stars with velocities consistent with the halo component than with either the KCC or GSS components, despite the fact that the velocity model for this field has the three components present in equal fractions. (The 50th percentile values for the fraction in each component are 33\% in the GSS, 32\% in the KCC, and 34\% in the halo; see Appendix~\ref{sec:app_velmodel}.) 
The numbers of stars with TiO features are roughly similar in the GSS and KCC\@.    
Because RGB stars with TiO features are cool, relatively metal-rich stars, the exclusion of these stars from the component-level abundance distribution functions is expected to bias the mean of the measured \feh\ distribution function (MDF) to more metal-poor values. Based on the relative number of TiO stars likely to belong to each component, this bias is expected to be somewhat larger for the halo MDF than the GSS and KCC MDFs. 

\begin{figure}[tb!]
\includegraphics[width=0.5\textwidth]{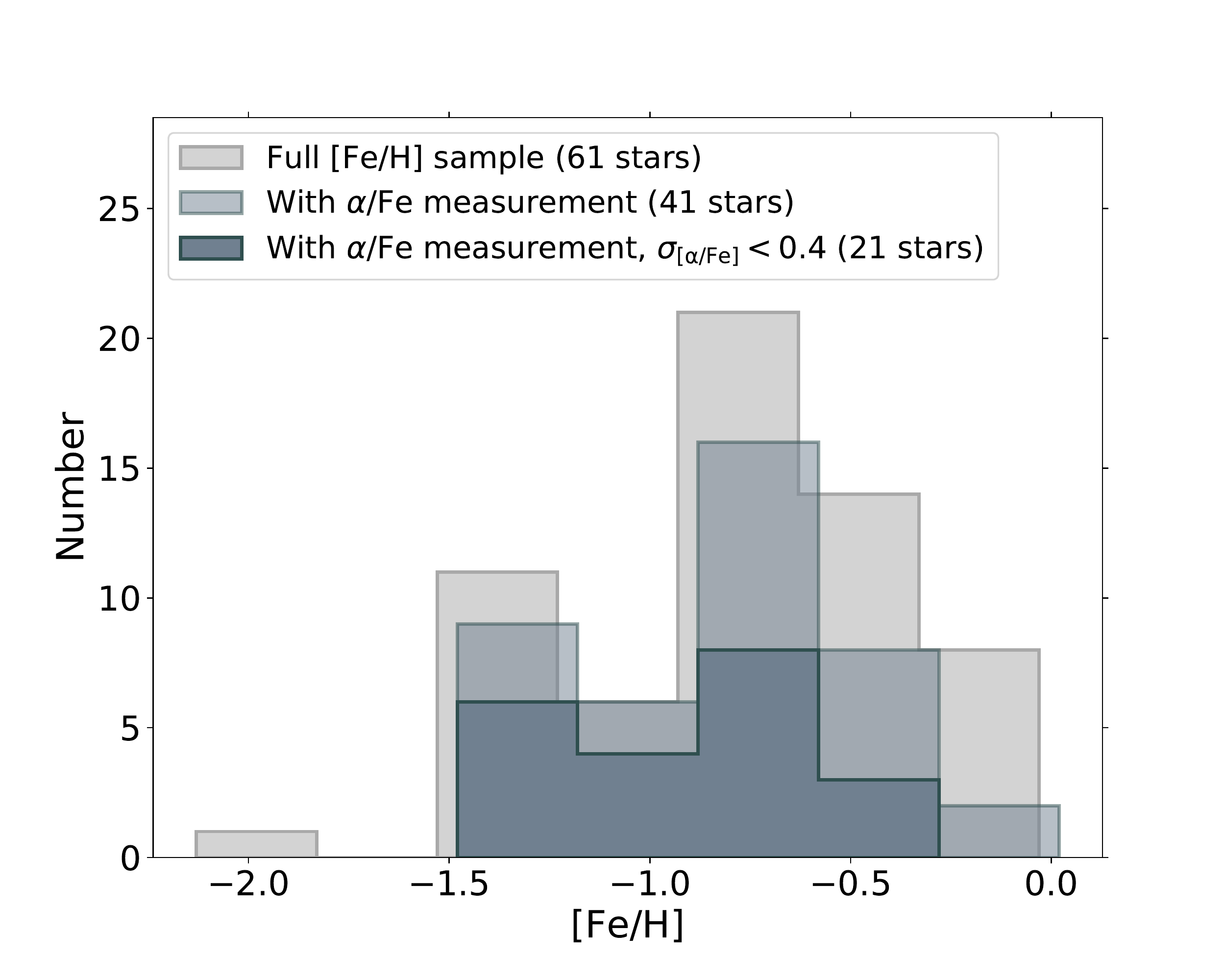}
\caption{
Distribution of \feh\ measurements that pass all criteria for inclusion in the final \feh\ sample (Section~\ref{sec:abund}). 
Cool RGB stars with evidence of TiO absorption in their spectra are excluded.   Stars that also have successful \afe\ measurements cover an equivalent range of \feh, with a similar distribution and fully consistent mean \feh\ values.  The \feh\ distribution of stars with \afe\ measurements that have uncertainties $<0.4$~dex is also similar, with a mean \feh\ that is only slightly (0.1~dex) more metal-poor. 
}
\label{fig:full_field_abunds_1d}
\end{figure}

Figure~\ref{fig:full_field_abunds_1d} shows the \feh\ distributions of stars with successful measurements.  Due to the number and strength of Fe lines in the available spectral regions, \feh\ abundances can be measured in a greater fraction of the spectroscopic sample than \afe.  The stars with \afe\ measurements span nearly the full range of measured \feh\ values, and there is no apparent bias in the \feh\ distribution of stars with successful \afe\ measurements if no uncertainty cut is made on the sample.    
The mean 
values of the distributions are $\langle$\feh\,$\rangle = -0.8\pm0.05$ for all stars with \feh\ measurements 
as well as for stars that also have \afe\ measurements.
If only stars with \afe\ uncertainties $<0.4$~dex are considered, $\langle$\feh\,$\rangle = -0.9\pm0.07$, which is consistent with the mean of the full \feh\ sample at the $\sim 1\sigma$ level. 
We include only \feh\ and \afe\ measurements with uncertainties $<0.4$~dex in the computation of the relevant distribution functions.  This does not remove any further \feh\ measurements, but does remove additional \afe\ measurements, resulting in 21 stars with \afe\ measurements passing all quality criteria.

\begin{figure}[tb!]
\includegraphics[width=0.5\textwidth]{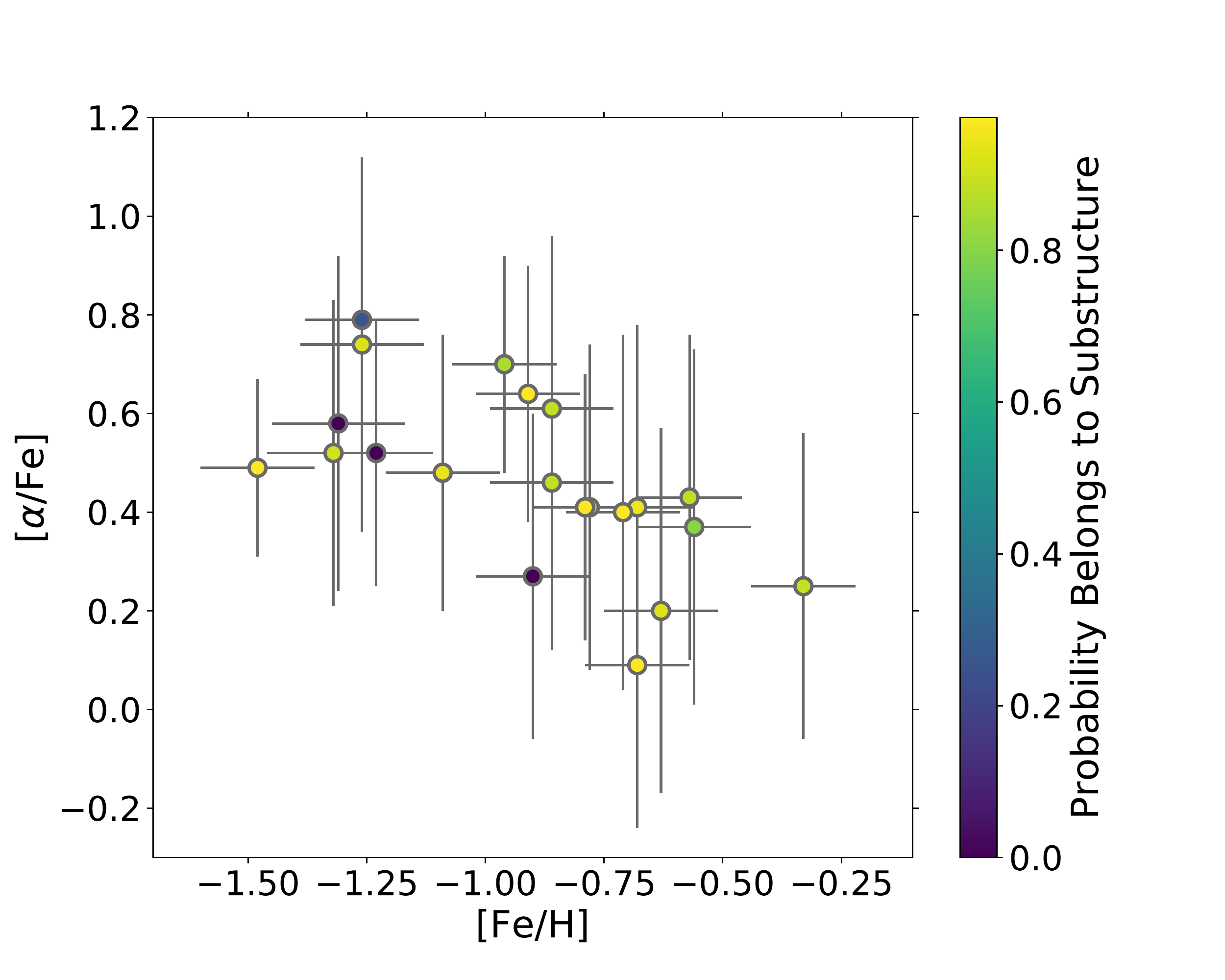}
\caption{Distribution of
\afe\ as a function of \feh\ for M31 RGB stars in the field with successful abundance measurements.  
As in Figure~\ref{fig:data_feh_vs_vel}, points are color-coded by the probability that they belong to one of the two tidal debris features identified in the field.
The stellar population in this field is significantly $\alpha$-enhanced at \feh\,$\lesssim -0.9$~dex, with decreasing $\alpha$-enhancement with increasing \feh\ above \feh\,$\gtrsim -0.9~$dex.  
}
\label{fig:full_field_abunds_2d}
\end{figure}

Figure~\ref{fig:full_field_abunds_2d} displays \afe\ as a function of \feh\ for stars with \afe\ measurements passing all quality criteria, including \afe\ uncertainties $\sigma$(\afe\,)$<0.4$~dex.  Intriguingly, the final sample of stars with \feh\ and \afe\ measurements show the ``knee''  feature characteristic of stellar populations that have extended star formation histories, continuing to form stars out of gas that has been polluted by the Fe-rich, $\alpha$-poor ejecta of Type~Ia SNe.  Furthermore, the location of the knee at the relatively metal-rich value of  \feh\,$\sim -0.9$~dex implies the stellar population self-enriched relatively rapidly before the onset of Type~Ia SNe.  This is consistent with the interpretation that a significant fraction of the stars in this field were deposited by a fairly massive progenitor that merged with M31 only within the last several Gyr \citep[e.g.,][]{fardal2013,hammer2018,dsouza2018}.  The abundance measurements with uncertainties greater than $0.4$~dex, which pass all other sample selection criteria, are largely consistent in the \afe\ vs.\ \feh\ plane with the measurements that pass the uncertainty cut.  We compare the \afe\ vs.\ \feh\ distribution of stars in this field within the broader context of M31 abundances, and discuss the implications for different merger scenarios, in Sections~\ref{sec:context} and \ref{sec:disc}.

Table~\ref{table_abund} 
presents the stellar parameters from the spectral synthesis fitting for the final sample of stars with \feh\ and \afe\ measurements.  These stars have passed all quality criteria, including a well constrained minimum in the $\chi^2$ contours for each of the parameters $T_{\rm eff}$, \feh, and (for \afe\ measurements) \afe, uncertainties in the abundance measurements of $<0.4$~dex, and no apparent TiO absorption in their spectra.

\begin{deluxetable*}{rllrrrrrrrrr}[htb!]
\tablecolumns{12}
\tablewidth{0pc}
\tablecaption{Parameters of Stars with Abundance Measurements.\label{table_abund}\tablenotemark{a,b}}
\tablehead{
\multicolumn{1}{c}{Object} & 
\multicolumn{2}{c}{Sky Coordinates} & 
\multicolumn{1}{c}{SN} &  
\multicolumn{1}{c}{Velocity} &  
\multicolumn{1}{c}{$T_{\rm eff}$} & 
\multicolumn{1}{c}{$\sigma(T_{\rm eff}$)} &
\multicolumn{1}{c}{log $g$} &
\multicolumn{1}{c}{[Fe/H]} &
\multicolumn{1}{c}{$\sigma$([Fe/H])} &
\multicolumn{1}{c}{[$\alpha$/Fe]} &
\multicolumn{1}{c}{$\sigma$([$\alpha$/Fe])} \\
\multicolumn{1}{c}{Name} &
\multicolumn{1}{c}{RA} & \multicolumn{1}{c}{Dec} & 
\multicolumn{1}{c}{(\AA\,$^{-1}$)} &  
\multicolumn{1}{c}{(km s$^{-1}$)} &  
\multicolumn{1}{c}{(K)} & 
\multicolumn{1}{c}{(K)} &
\multicolumn{1}{c}{(dex)} &
\multicolumn{1}{c}{(dex)} &
\multicolumn{1}{c}{(dex)} &
\multicolumn{1}{c}{(dex)} &
\multicolumn{1}{c}{(dex)}
} 
\startdata
2071585 & 00h43m28.88s & +40d03m41.3s &   9.39 & $-$436.2 &     4875 &     71 &   1.40 &  $-$0.71 &   0.14 & ... & ... \\
2071602 & 00h43m33.07s & +40d03m40.7s &   6.51 & $-$423.6 &     4484 &     78 &   1.36 &  $-$0.47 &   0.15 & ...  & ... \\
2071604 & 00h43m31.69s & +40d03m41.1s &   8.91 & $-$427.3 &     3855 &     47 &   1.26 &  $-$0.09 &   0.13 & ... & ... \\
2071607 & 00h43m27.78s & +40d03m39.0s &   5.66 & $-$443.0 &     4403 &     69 &   1.31 &  $-$0.08 &   0.14 & ... & ... \\
2071820 & 00h43m40.93s & +40d03m15.1s &   8.55 & $-$277.6 &     4484 &     68 &   1.32 &  $-$1.39 &   0.22 & ... & ... \\
2071869 & 00h43m35.74s & +40d03m04.7s &  15.85 & $-$378.5 &     4087 &     23 &   0.59 &  $-$1.26 &   0.12 &   0.79 &   0.25 \\
2072033 & 00h43m27.02s & +40d02m39.3s &  14.43 & $-$427.9 &     3841 &     21 &   0.81 &  $-$0.86 &   0.13 &   0.61 &   0.35 \\
2072073 & 00h43m37.82s & +40d02m35.7s &  14.14 & $-$459.4 &     3807 &     18 &   0.72 &  $-$0.56 &   0.12 &   0.37 &   0.36 \\
2072102 & 00h43m44.90s & +40d02m30.2s &  12.75 & $-$320.9 &     4437 &     38 &   1.08 &  $-$0.90 &   0.12 &   0.27 &   0.33 \\
2072184 & 00h43m38.62s & +40d02m19.1s &  13.41 & $-$508.3 &     3961 &     20 &   0.70 &  $-$0.68 &   0.12 &   0.41 &   0.37 \\
\enddata
\tablenotetext{a}{Measurements of \feh\ and \afe\ included here pass all quality criteria discussed in Section~\ref{sec:abund}.  Stars with evidence of TiO in their spectra have been excluded from the final sample of abundances.
}
\tablecomments{This table is provided as an example of the full table, which will be published in its entirety in machine readable format.
}
\end{deluxetable*}

\section{Abundance Distributions of the Individual M31 Components}\label{sec:mdfs}

A clean separation of the three previously identified kinematical components among the M31 stars in this field cannot be achieved using simple cuts on the observed line-of-sight velocities (Figure~\ref{fig:vel_model}).   
Instead, we derive probabilistic
abundance distributions based on the probability that a given star belongs to each component.  This maximizes completeness while minimizing contamination in the resulting component-level abundance distribution functions.  We describe the computation of the probabilistic abundance distributions in Section~\ref{sec:compute_prob_mdfs}, assess the impact of uncertainties in the velocity model in 
Section~\ref{sec:vel_model_uncertainties}, and discuss sources of systematic bias in the MDFs in Section~\ref{sec:mdf_biases}.

\subsection{Computation of Probabilistic Abundance Distributions}\label{sec:compute_prob_mdfs}
The probabilistic distribution functions (DFs) are computed as the sum of normalized Gaussians, $\mathcal{G}$, 
with the mean given by a star's abundance measurement ($\mu_i$) and the standard deviation given by the uncertainty in the 
abundance measurement ($\sigma_i$).  The contribution of each star, $\mathcal{G}_i$, to the sum for a given component $j$ is weighted by the probability that 
the star belongs to that component, $p_{i, j}$.  Finally, the summed $\mathcal{G}_i$s are divided by the sum of the probabilities, resulting in a normalized DF for component $j$:
\begin{equation}\label{eqn:df}
\mathcal{D}_{j}= \frac{1}{\sum_{i=1}^{N_{\rm stars}} p_{i, j}}\sum_{i=1}^{N_{\rm stars}} p_{i, j}\,\mathcal{G}(\mu_i,\sigma_i)
\end{equation}

\begin{figure}[tb!]
\includegraphics[width=0.5\textwidth]{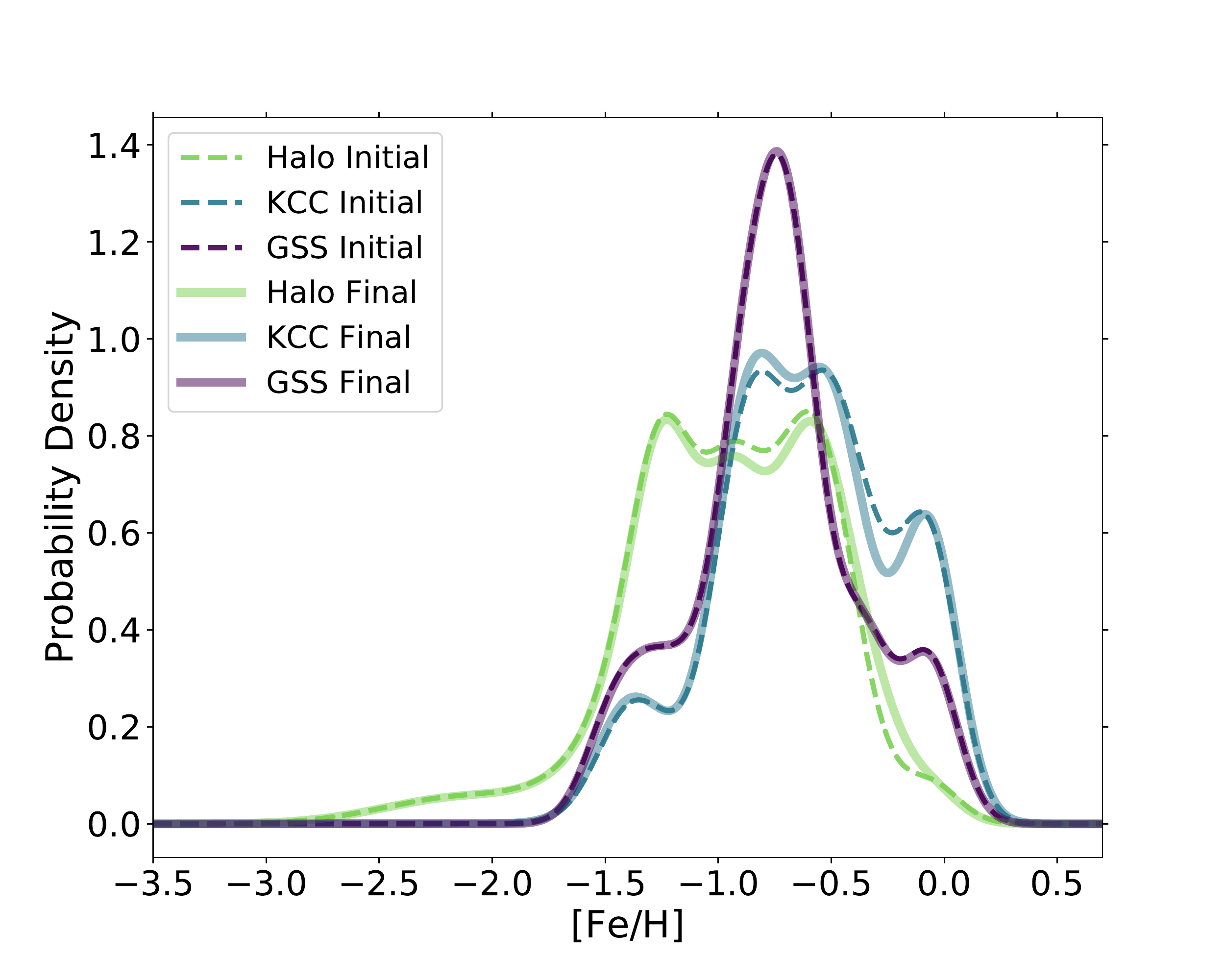}
\caption{
Probabilistic MDFs 
for each of the three M31 components present in the field, derived using the 50th-percentile parameter values of the velocity model  (Section~\ref{sec:compute_prob_mdfs}).    
Dashed curves show MDFs produced by weighting each
stars' contribution by the probability the star belongs to each of the three components based solely on its velocity.
Solid curves show the MDFs produced when including previous iterations of the MDF of each component to compute
the joint probability the star belongs to each of the three components, including both the star's velocity and [Fe/H].  
In general, iterations including the MDF affect primarily the halo and KCC component, which is expected based 
on the greater overlap between these two components in the velocity model.  However, even for these two components, 
including MDFs from the prior iteration 
in the calculation of the probabilities has only a mild effect on the final component DFs.  
}
\label{fig:mdfs_iter_example}
\end{figure}

For each star, the observed line-of-sight velocity is used to compute initial 
probabilities, $p_{i, j}^{\rm init}$ that it belongs to the GSS, 
the KCC, and the kinematically hot halo 
using the velocity model of \citet{gilbert2018} (Appendix~\ref{sec:app_velmodel}).
These probabilities are used to calculate an initial probabilistic MDF for each component, $\mathcal{D}_{j}^{\rm init}$(\feh) 
(dashed curves, Figure~\ref{fig:mdfs_iter_example}).

However, given the overlap in velocity space of the M31 components in this field (Figure~\ref{fig:vel_model}), a given star may be
more probable to belong to one component based on velocity alone while in reality it belongs to a 
different component.  
Thus, we utilize the \feh\ measurements to refine the abundance DFs for 
each component.    

After computing the initial MDFs [$\mathcal{D}_{j}^{\rm init}$(\feh)] for each
component based on probabilities derived solely from line-of-sight velocity [$p_{i, j}^{\rm init}(v_i)$], 
we use these MDFs in conjunction
with the velocity model to compute the joint probability that star $i$, with velocity  $v_i$ and metallicity \feh$_i$, was drawn from component $j$:   
\begin{equation}
p_{i, j}(v_i, {\rm [Fe/H]}_i) = \frac{p_{i, j}(v_i)\,p_{i, j}({\rm [Fe/H]}_i)}{\sum_{k=1}^3 {p_{i, k}(v_i)\,p_{i, k}({\rm [Fe/H]}_i)}},
\label{eqn:mixingfracs}
\end{equation}
where the normalization term enforces the assumption that the star is drawn from one of the three components.  In Equation~\ref{eqn:mixingfracs}, $p_{i, j}(v_i)$ is computed from the M31 halo velocity model evaluated at $v_i$ \citep[Appendix~\ref{sec:app_velmodel} and][]{gilbert2018}:
\begin{equation}
p_{i, j}(v_i) = \frac{f_j\,\mathcal{G}(v_i | \mu_j, \sigma_j)}{\sum_{k=1}^3 f_k\,\mathcal{G}(v_i | \mu_k, \sigma_k)},
\end{equation}
where $\mu_j$ is the mean velocity, $\sigma_j$ is the velocity dispersion, and $f_j$ is the fraction of stars belonging to the $j$th component (constrained to $\sum_{k=1}^3 f_k = 1$).
The \feh-based probability in Equation~\ref{eqn:mixingfracs}, $p_{i, j}({\rm [Fe/H]}_i)$, is given by the previously computed $\mathcal{D}_{j}$ evaluated at \feh$_i$:
\begin{equation}
p_{i, j}({\rm [Fe/H]}_i) = \frac{\mathcal{D}_{j}({\rm [Fe/H]}_i)}{\sum_{k=1}^3 \mathcal{D}_{k}({\rm [Fe/H]}_i)}
\end{equation}

The joint probabilities from Equation~\ref{eqn:mixingfracs} are used in Equation~\ref{eqn:df} to compute a new MDF.
This process is then iterated, with 
MDFs ($\mathcal{D}_{j}$(\feh)) computed in the previous step 
used to compute a new joint probability for each star,  
and the new joint probabilities used to compute updated MDFs.  
This process was continued until the joint probabilities, $p_{i, j}(v_i,\feh_i)$, 
converged.  Convergence was declared when all $p_{i, j}(v_i,\feh_i)$ 
changed by less than 0.02\% from the previous iteration to the current iteration. 
This criterion was found to be more than sufficient to achieve stability in the calculated MDFs.  

Figure~\ref{fig:mdfs_iter_example} displays the probabilistic component-level \feh\ 
DFs based on the initial velocity-based $p_{i, j}^{\rm init}(v_i)$ and final joint $p_{i, j}^{\rm final}(v_i,\feh_i)$ probabilities, calculated using the 50th percentile values for the velocity model parameters.  The differences between the initial [$\mathcal{D}_{j}^{\rm init}$(\feh), computed using only $p_{i,j}(v_i)$] and final DFs are small.  This is expected, as Figure~\ref{fig:data_feh_vs_vel} indicates there are relatively small differences in the range and mean values of \feh\ between the three components in the field.  Although there is considerable overlap in the \feh\ distributions, the halo component includes a larger fraction of metal-poor stars than the GSS or KCC components, while the KCC component appears to be slightly more metal-rich than the GSS\@.  The 16th, 50th, and 84th percentiles of the cumulative \feh\ probabilistic DF for each component are reported in Table~\ref{components_table}.

Finally, we use the converged joint probabilities, $p_{i, j}^{\rm final}$, for every star with an \afe\ measurement passing all quality criteria (Section~\ref{sec:abund}) 
to produce a probabilistic representation of the \afe\ vs.\ \feh\ distribution for each component.  In analogy with the procedure used for the 1-dimensional DFs, a single star's contribution to a component's 2-dimensional DF is represented as a 2-dimensional Gaussian, with mean and sigmas given by the \afe\ and \feh\ measurements and measurement uncertainties, assuming no covariance, and with a weight proportional to its probability of belonging to that component. The total DF is the sum over all stars with \afe\ measurements.  

\begin{figure}[h!]
\plotone{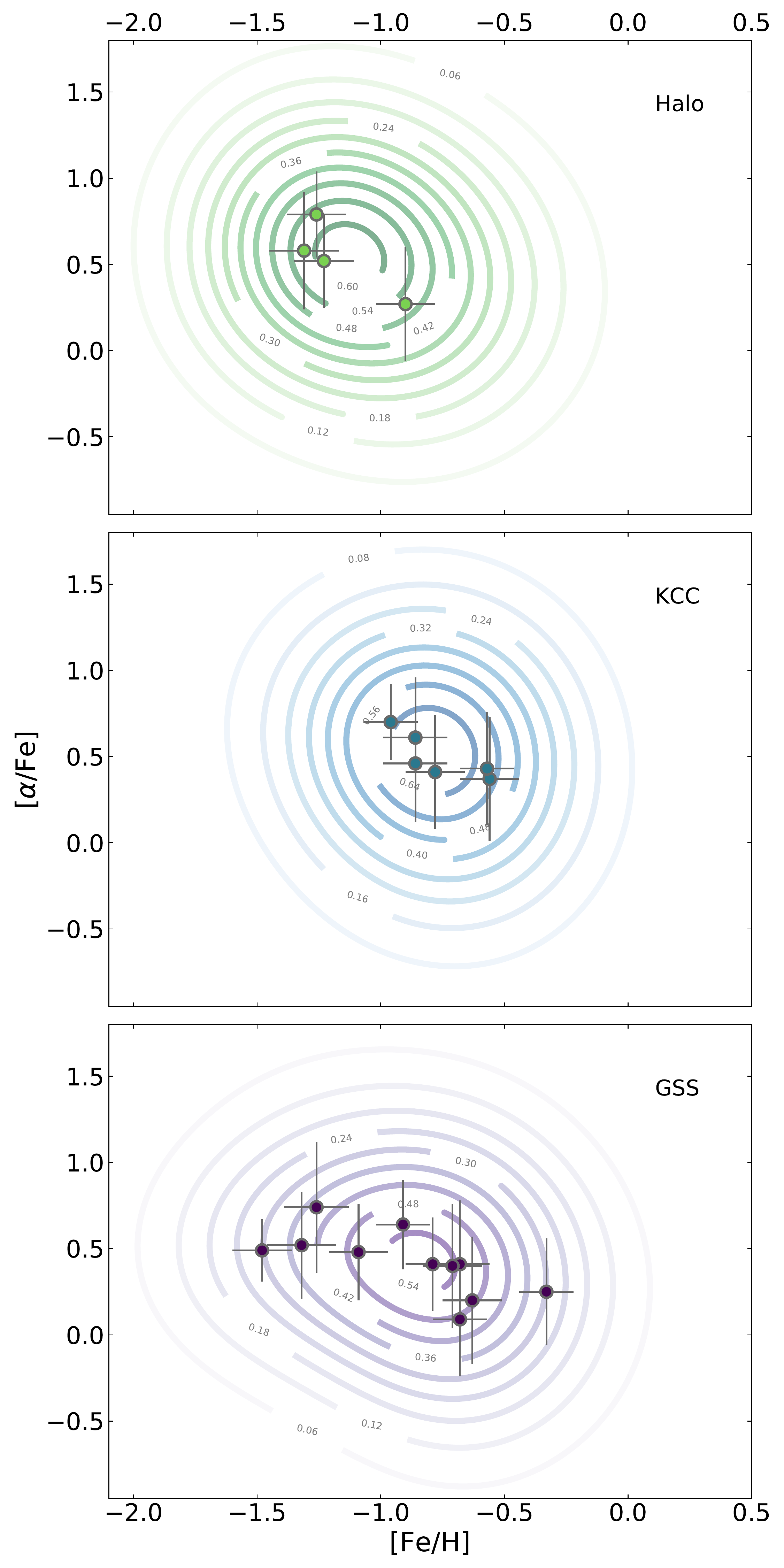}
\caption{
Probabilistic \afe\ vs.\ \feh\ distributions (contours; Section~\ref{sec:compute_prob_mdfs}) for each of the three components present in the GSS field, computed using the probability of membership in a given component 
for all stars with \afe\ measurements 
passing the quality criteria (Section~\ref{sec:abund}).  
To provide an indication of the number of stars likely associated with each component, 
individual data points are overlaid on the probabilistic contours. The GSS and KCC panels show individual measurements for stars within $\pm 2\sigma_v$ of the mean velocity of the component. 
The remainder of the stars are shown in the Halo panel. 
The GSS component has 
enhanced \afe\ at low \feh\, and decreasing $\alpha$-enhancement with increasing \feh\ for \feh\,$\gtrsim -0.9$~dex.
}
\label{fig:afe_components}
\end{figure}

Figure~\ref{fig:afe_components} shows the probabilistic \afe\ vs.\ \feh\ contours,  
separated into kinematic components using the 50th-percentile parameter values from the velocity model (Appendix~\ref{sec:app_velmodel} and Figure~\ref{fig:corner}).  Also plotted for comparison are stars within $\pm 2\sigma_v$ of the mean velocity of the GSS and KCC component; the halo panel shows stars that have velocities further removed than $2\sigma_v$ from both the GSS and KCC components.  The individual measurements are included in Figure~\ref{fig:afe_components} in order to give a sense of the relative size of the stellar sample that contributes to each two-dimensional DF.  

The GSS component has the largest number of stars with both high probabilities of component membership as well as valid \feh\ and \afe\ measurements.  Both the likely GSS stars and the probabilistic two-dimensional DF are consistent with a relatively flat $\alpha$-enhancement until \feh\,$\sim -0.9$~dex, and with decreasing $\alpha$-enhancement with increasing \feh\ for \feh\,$\gtrsim -0.9$~dex.
While the current sample sizes are too small to yield specific insights for the halo and KCC component, we note that the two-dimensional KCC DF and the stars with high probabilities of belonging to the KCC are both fully consistent with the \afe\ vs.\ \feh\ distribution of GSS stars.    

\subsection{Assessment of the Impact of Uncertainties in the Velocity Model}\label{sec:vel_model_uncertainties}

\begin{figure*}[tb!]
\includegraphics[width=0.5\textwidth]{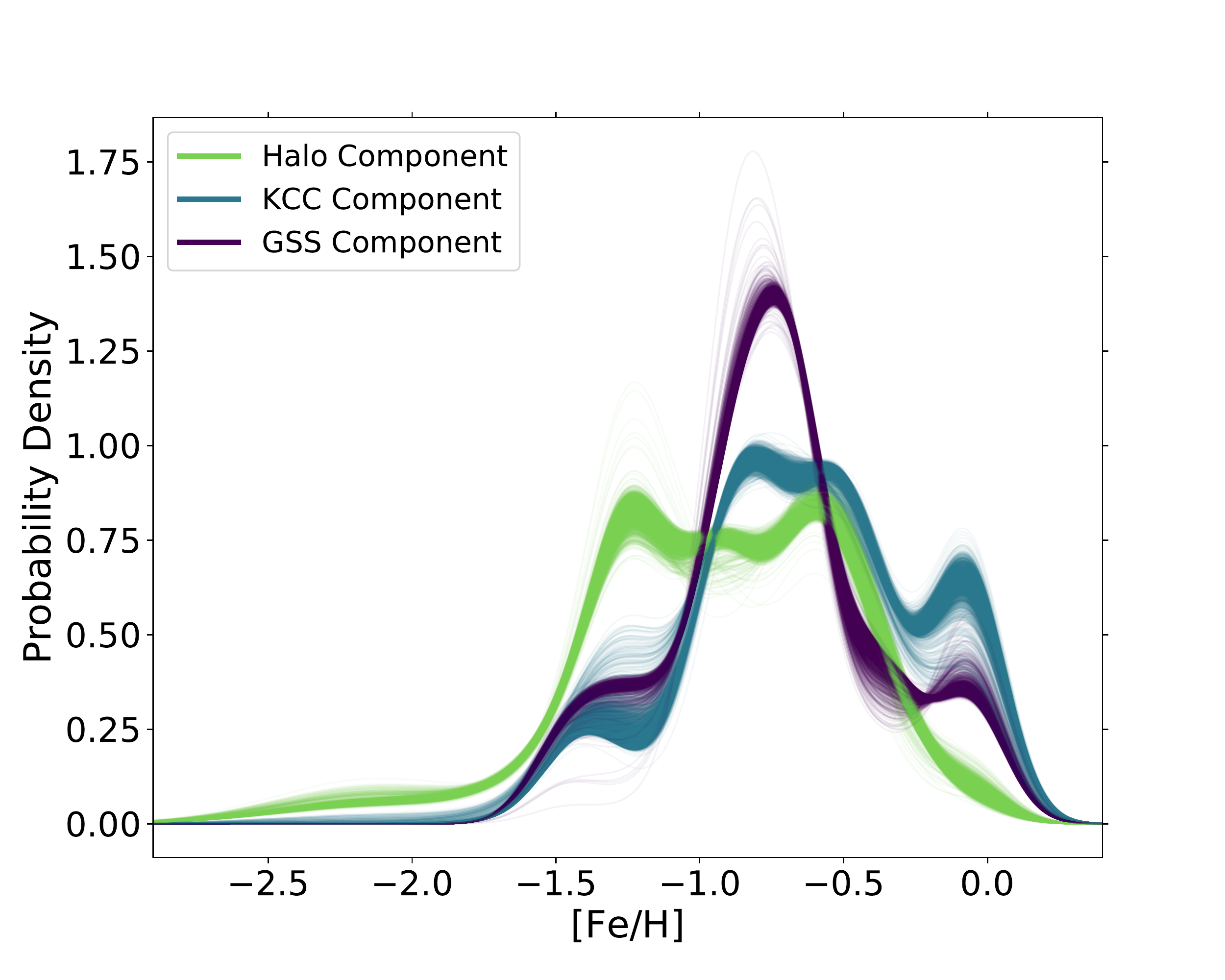}
\includegraphics[width=0.5\textwidth]{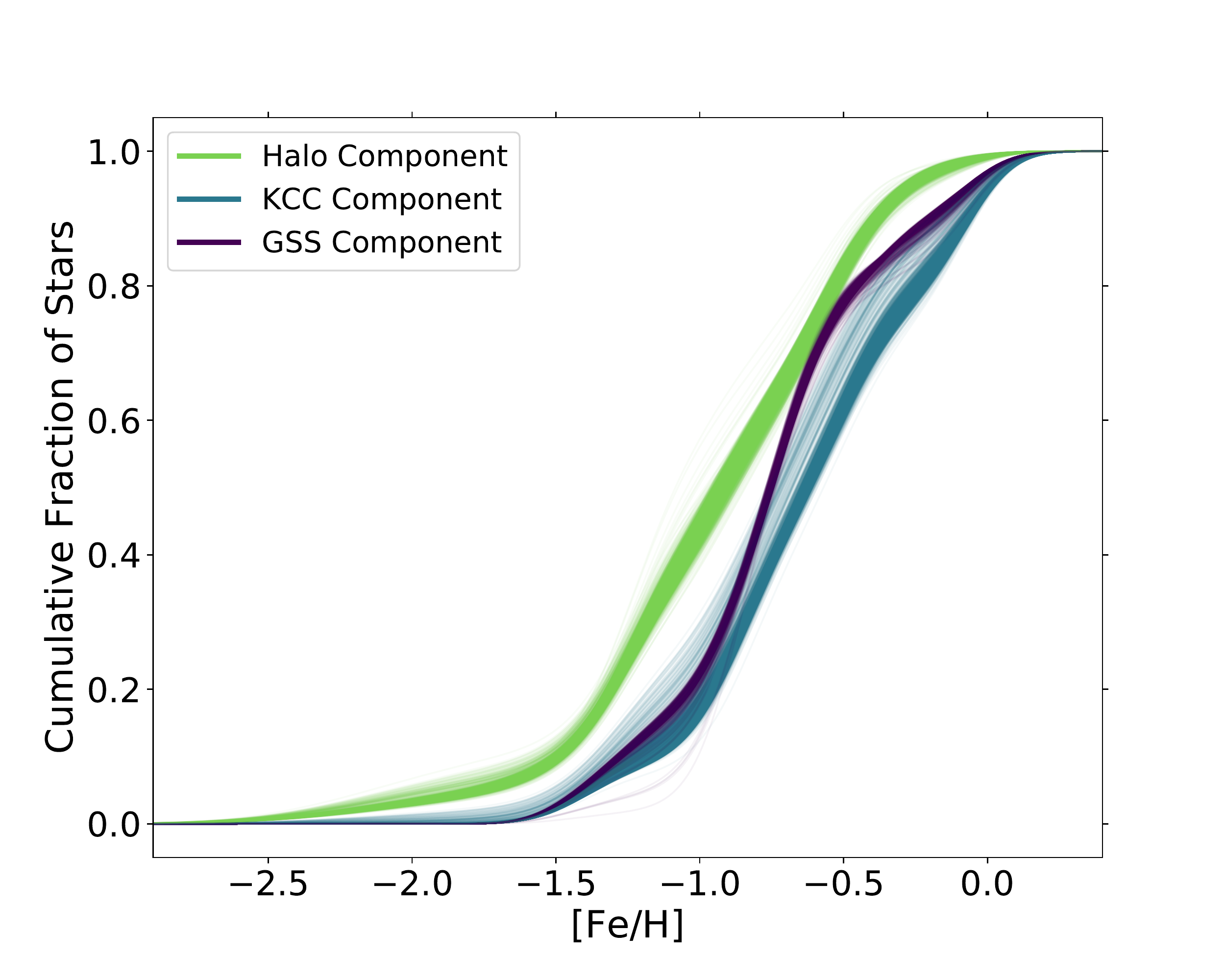}
\caption{
Probabilistic \feh\ DFs in normal ({\it left}) and cumulative ({\it right}) form for each of the three components in the field (Section~\ref{sec:vel_model_uncertainties}).  The probability that a given star belongs to a given component was computed using velocity model parameters from 2000 random draws of the MCMC chains published by \citet{gilbert2018} (Appendix~\ref{sec:app_velmodel}).  The range of the component-level DFs encapsulates the underlying uncertainties in our knowledge of the velocity distribution of M31 stars in this field.  The MDFs of the GSS and KCC have a substantial degree of overlap. The halo component, while also containing a majority of stars with \feh\,$> -1.0$~dex, is on average more metal-poor than either the GSS or KCC\@.
}
\label{fig:mdfs}
\end{figure*}

The previous section presented the probabilistic component-level DFs for \feh\  and \afe\  using the 50th-percentile values of the marginalized one-dimensional posterior probability distribution functions for each of the parameters of the \citet{gilbert2018} velocity model for this field.  However, analyzing these DFs alone ignores the known uncertainties and covariances in the velocity model parameters (Appendix~\ref{sec:app_velmodel}, Figure~\ref{fig:corner}).  

We therefore repeated the calculation of component-level MDFs described above, using different values for the parameters in the velocity model.
We made 2000 random draws from
the MCMC chains published in \citet{gilbert2018}.  Each draw yields a set of parameters for the mean velocities, velocity dispersions, and mixing fractions of each 
component in the field.  The analysis described in  Section~\ref{sec:compute_prob_mdfs} was performed for each draw.  This utilized the full posterior probability distributions computed by \citet{gilbert2018} to estimate the variance in the component-level DFs caused by uncertainties in the underlying velocity model for this field.

The resulting DFs  are shown in Figure~\ref{fig:mdfs}. 
The range of DFs for each component reflects the impact of uncertainties in the velocity model parameters on the 
probability that star $i$ is drawn from component $j$.  This confirms that the GSS and KCC MDFs are quite similar, with substantial overlap between the two MDFs.  The subtle differences between the GSS and KCC MDFs are in fact qualitatively similar to what was seen in the comparison of the photometrically derived MDFs of the GSS and the KCC presented by \citet[][their Figure~17]{gilbert2009gss}, with a slight overabundance of stars at the metal-rich end (\feh\,$\gtrsim -0.7$~dex) in the KCC component compared to the GSS component.  While the halo MDF is on average more metal-poor than either the GSS or KCC MDFs,  \feh\,$>-1.0$~dex for approximately half of the stars in the halo component.

\subsection{Sources of Bias in the Observed MDFs}\label{sec:mdf_biases}
Observed MDFs of stars near the tip of the RGB are prone to a systematic bias against the recovery of \feh\ measurements for metal-rich stars.  This is due in part to the inherently redder colors and fainter $I$-band magnitudes of metal-rich stars, leading to a lower spectral SNR\@.  This is shown by the isochrones in Figure~\ref{fig:cmd}.  For the most metal-rich isochrones, the tip of the RGB is a half magnitude to a magnitude fainter than the tip of the metal-poor RGB\@.  

This effect is partially mitigated by the strength of the Fe lines in metal-rich stars.   Moreover, our spectroscopic targets lie well below the tip of the RGB, with a significant fraction of the targets below the magnitude at which the `turnover' in the metal-rich isochrones occur (Figure~\ref{fig:cmd}).    
Nevertheless, Figure~\ref{fig:cmd} shows that the spectral synthesis-based \feh\ measurements failed for the reddest of the spectroscopic targets.  However, similar numbers of stars with very red colors have failed \feh\ measurements across the three components.  Of stars with failed \feh\ measurements and \fehp\,$>-0.5$, three have velocities consistent with the halo component, four with the GSS, and five with the KCC (Figure~\ref{fig:fehphot_vs_vel}).  Since the MDFs (as well as the CMD distribution) of stars in all three components are fairly similar, the magnitude of the bias introduced by the failure to recover \feh\ for the reddest stars is likely to be similar across all three components, and thus is unlikely to significantly affect the relative comparison of the component-level MDFs.

However, the exclusion of cool, metal-rich stars with TiO features in their spectra results in an additional bias against recovery of the metal-rich end of the MDFs.  Moreover, exclusion of these stars will have a larger effect on the halo MDF than the GSS and KCC MDFs.  While the three components comprise similar fractions of the stellar population in the field, there are approximately twice 
as many stars with TiO absorption and velocities that are consistent with the halo component (14 stars) than with either the GSS (8 stars) or the KCC (7 stars) components (Figure~\ref{fig:data_feh_vs_vel}).  Thus, the true halo MDF may in fact have more metal-rich stars than indicated by Figure~\ref{fig:mdfs}.

Finally, we note that there are additional biases, unrelated to observational completeness effects, in MDFs derived from the RGB.  \citet{manning2017} demonstrated that metal-rich intermediate-age to old stars ($\gtrsim 6$~Gyr) are underrepresented relative to their true densities in the top two magnitudes of the RGB\@. The magnitude of this bias is dependent on the star formation history (SFH) and chemical evolution of the population.  If the SFH of the GSS, KCC, and halo are fairly similar, this will have little impact on the relative comparison of the MDFs.  For a purely old ($\gtrsim 10$~Gyr) stellar population, this bias is negligible. Therefore if the underlying halo is purely old \citep[while the GSS and KCC are not;][]{brown2006}, the difference in the true underlying MDFs between the halo and substructure in this field is likely larger than observed.  However, existing SFHs indicate that M31's halo is likely more complex than a monolithically old population \citep{brown2008}. 

\section{The Abundances of the GSS Field in Context}\label{sec:context}

In this section, we compare the \feh\ and \afe\ abundances of stars in the GSS field with the \feh\ and \afe\ abundances of stars in M31's dSphs and elsewhere in M31's stellar halo (Figure~\ref{fig:roadmap}).  We also compare the GSS field abundances to those of stars in the progenitor of the MW's largest tidal stream, Sagittarius.

\begin{figure*}[tb!]
\includegraphics[trim=0cm 2cm 2cm 4cm,clip,width=\textwidth]{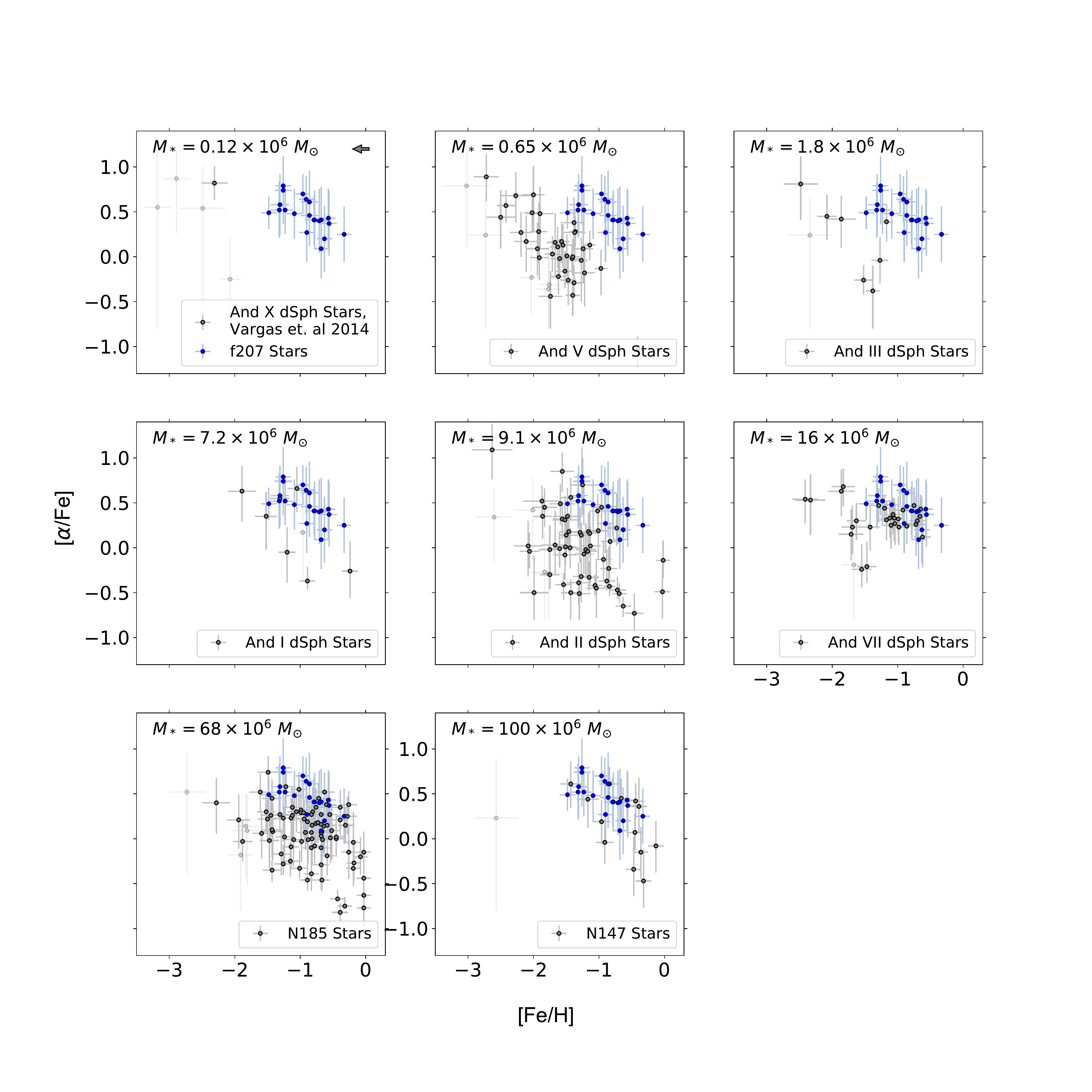}
\caption{
Comparison of the \afe\ vs. \feh\ distribution of stars in the GSS field with stars in M31 dwarf satellite galaxies \citep{vargas2014} covering a range of stellar mass and metallicity (Section~\ref{sec:context}).  Transparent points show measurements from \citeauthor{vargas2014} with a large mean uncertainty ($\ge 0.4$) in \afe.  
The stellar masses of the dwarf satellites listed in each panel are computed using luminosity estimates by \citet[And I, III , V, VII, and X]{tollerud2012} 
or compiled by \citet[And II, N147, and N185]{mcconnachie2012} 
and assume the stellar mass-to-light ratios compiled by \citet{woo2008}. 
The arrow in the And~X panel shows the typical offset in \feh\ measurements between \citet{vargas2014} and our measurements; \citeauthor{vargas2014}'s \feh\ measurements are systematically more metal-rich.     
The abundances of stars in the GSS field are most similar to the more massive of M31's surviving dwarf satellite galaxies, implying that the tidal debris in this field was deposited by a fairly massive progenitor.  The higher average $\alpha$-enhancement of the GSS indicates it also had a higher efficiency of star formation.  
}
\label{fig:f207_vs_m31dwarfs}
\end{figure*}

Figure~\ref{fig:f207_vs_m31dwarfs} compares the \afe\ and \feh\ abundances of stars in the GSS field with abundances of stars belonging to M31 dwarf satellites \citep{vargas2014}, ordered by increasing stellar mass.  
We have compared 
our measurements to those published by \citet{vargas2014} using a set of M31 dSph stars observed in both surveys (E.\ Kirby et al., in prep.).  While we find a small offset in \feh\ between \citeauthor{vargas2014}'s and our measurements (\feh$_{\rm V14} - $ \feh$_{\rm Kirby} \sim 0.2-0.3$~dex, indicated in Figure~\ref{fig:f207_vs_m31dwarfs}), the \afe\ measurements are consistent within the uncertainties.  Thus, the true differences in \feh\ between the GSS field and M31's dSphs are slightly larger than indicated in Figure~\ref{fig:f207_vs_m31dwarfs}.

The abundances of the stars in the GSS field are most comparable to the stellar abundances of the most massive satellites in the sample (NGC~147 and NGC~185).  The GSS field is significantly more metal-rich than the less-massive M31 dSphs.  This is fully consistent with the interpretation that the progenitor of the GSS was a relatively massive satellite 
that was tidally disrupted during a recent encounter with M31 (Section~\ref{sec:disc}).
Stars in the GSS field are also on average more $\alpha$-enhanced than the surviving M31 dwarf satellites for which there are abundance measurements.  Therefore, the progenitor of the GSS reached a higher metallicity than most surviving satellites by the time Type~Ia supernovae began to explode.  This indicates that it experienced a higher efficiency of star formation (mass of stars formed per mass of gas).

\begin{figure}[tb!]
\includegraphics[width=0.5\textwidth]{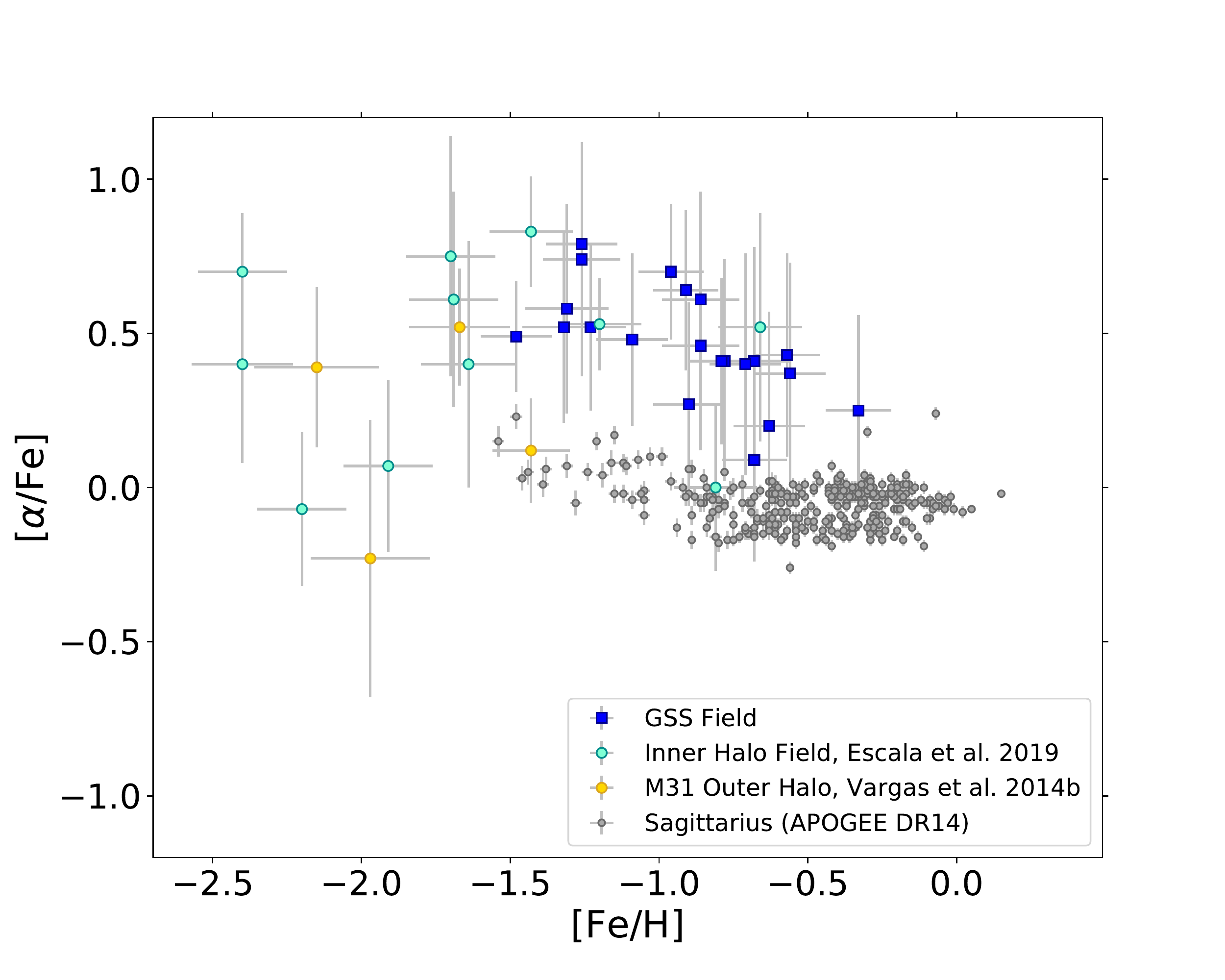}
\caption{
Comparison of the \afe\ and \feh\ abundances of stars in the GSS field with the abundances of stars in other M31 halo fields \citep{vargas2014apjl, escala2019}, and stars in the core of the Sagittarius dSph \citep[using the selection criteria of][]{hasselquist2017}, which is the progenitor of the MW's largest stellar tidal stream (Section~\ref{sec:context}).   Of the comparison samples shown here, the abundances of stars in the GSS field are most similar to the abundances of stars in the inner M31 halo field, which is near M31's southeast minor axis and does not appear to be dominated by pollution from a recent merger event.  However, the stars in the GSS field are on average more metal-rich than stars in the inner halo minor axis field.  The progenitor of the Sagittarius stream appears to have experienced a very different star formation history than the progenitor of the GSS\@.   
}
\label{fig:f207_vs_m31halo}
\end{figure}

Figure~\ref{fig:f207_vs_m31halo} compares the \afe\ and \feh\ abundances of stars in the GSS field with abundances of M31 halo stars in other fields, as well as stars in the core of the Sagittarius dSph.  
Existing M31 halo star measurements prior to our survey consisted of four stars in M31's outer halo with abundances measured by \citet{vargas2014apjl}, which were identified in three dSph fields (And~III, And~XIII, and And~II) at distances from M31's center of \rproj $ = 69$, 117, and 144~kpc, respectively.  
These stars were identified as halo stars, rather than dSph members, by virtue of being well removed from the locus of dSph member stars in either line of sight velocity or spatial position. Two of the outer halo stars have velocities consistent with that of their nearby dSph, but spatial positions that place them well beyond the dSph's tidal radius \citep{vargas2014apjl}.  

Also shown in Figure~\ref{fig:f207_vs_m31halo}  are abundances of 11 stars in an inner halo field (`f130,' Figure~\ref{fig:roadmap}) at \rproj\,$=22$~kpc on M31's minor axis, recently published by \citet{escala2019}.  This inner halo minor axis field lies several kpc in projection beyond the outermost edge of the GSS-related shell feature identified along the minor axis by \citet{gilbert2007}.  In contrast to the GSS field, stars in the inner halo minor axis field show no evidence of distinct tidal debris features.  They are well mixed in velocity vs.\ metallicity space, and they are consistent with being drawn from a single Gaussian with a relatively large velocity dispersion.  

On average, the stars in the GSS field are significantly more metal-rich than stars in either the inner halo minor axis field or M31's outer halo; this is true for any of the components in the field, including the halo component.  Stars in the GSS field and the inner halo minor axis field are on average more $\alpha$-enhanced than the M31 outer halo stars, with stars in the two inner halo fields having a similar range of \afe\ abundances. The existing measurements indicate a potential systematic difference between \afe\ vs.\ \feh\ abundance distributions of stars in M31's inner and outer halo. However, given the very small number of outer halo stars with existing measurements and the complexity of M31's inner halo, larger samples will be required to make quantitative statements about the trends of \afe\ and \feh\ with projected distance in M31's stellar halo. 

The Sagittarius sample is chosen using the selection criteria of \citet{majewski2013} and \citet{hasselquist2017}, which isolate RGB stars in the region of the Sagittarius core.  The abundances are drawn from APOGEE DR14 \citep{majewski2017}.  
The GSS field is significantly more alpha-enhanced 
than Sagittarius.  The difference in their \afe\ vs.\ \feh\ distributions indicate that the two populations experienced fundamentally different star formation histories.   
Sagittarius must have formed stars slowly over a timescale extended enough for Type~Ia supernovae to depress \afe\ to sub-solar values.  \citeauthor{hasselquist2017}\ interpreted some of the deficiency in $\alpha$-elements as evidence of a top-light initial mass function (IMF), but \citet{mucciarelli2017} argued that a normal IMF could also produce the observed \afe\ trend in Sagittarius.  IMF variations are unlikely to explain the steeply declining trend of \afe\ vs.\ \feh\ that we observe in the GSS\@.  Rather, the progenitor to the GSS seems to have experienced a rapid decline in star formation rate (SFR) at late times, allowing Type~Ia supernovae to overtake core collapse supernovae as the dominant sites of nucleosynthesis.  This is consistent with both the star formation history measured from deep HST imaging \citep{brown2006, brown2006a} and the merger scenarios posited as the origin of the GSS (Section~\ref{sec:disc}).  In comparison, the late-time evolution of Sagittarius seems to reflect a nearly steady state of star formation, where \afe\ is low but evolving very slowly.

Abundance measurements in the LMC and SMC show that these galaxies, while more $\alpha$-enhanced than Sagittarius, are also significantly less $\alpha$-enhanced than stars in the GSS field, with mean \afe\ abundances near 0~dex over a large range in metallicity, requiring very low star formation efficiencies \citep[e.g.,][]{pompeia2008,lapenna2012,vanderswaelmen2013,nidever2019}.  Recent work by \citet{nidever2019} constrains the ``$\alpha$-knee'' in both the LMC and SMC to \feh\,$\lesssim -2.2$, in stark contrast to the location of the ``$\alpha$-knee'' in the GSS field at \feh\,$\sim -0.9$ as well as the location of the ``$\alpha$-knee'' in other MW dwarf galaxies \citep[including at \afe\,$=-1.3$ in the Sgr stream][]{deboer2014}.  
\citet{nidever2019} postulate that this would be consistent with the MCs forming in a more isolated environment, and falling into the MW potential for the first time only recently \citep{besla2007,besla2012}.  The GSS progenitor, however, appears to have largely stopped forming stars by $\sim 6$~Gyr ago \citep{brown2006}, indicating that although it may have disrupted as recently as $\lesssim 1$~Gyr ago, it is likely to have been influenced by M31's grativational potential well before then.

\section{Discussion}\label{sec:disc}

\subsection{Implications for the Minor vs.\ Major Merger Scenarios}
High-resolution numerical simulations of the formation of the GSS through a minor merger have been extremely successful in reproducing the observed stellar density and line-of-sight stellar velocity distributions in M31's GSS and have demonstrated that a series of shells in M31's inner regions is in fact the continuation of the GSS\@. These models indicate the GSS progenitor had an initial stellar mass of 1\,--\,$5\times10^9$\Msun\ (placing it in a comparable range of stellar mass as that of the LMC and M33) and experienced a disruptive pericentric passage with M31 $\lesssim 1$~Gyr ago \citep{fardal2006, fardal2007, fardal2008, fardal2012, fardal2013, mori2008, sadoun2014, kirihara2014, kirihara2017}. 

However, morphological differences between M31 and the MW  \citep[see][and references therein]{hammer2018}, as well as the recent discoveries of a rotating inner halo \citep{dorman2012}, a population of heated disk stars with halo-like kinematics \citep{dorman2013}, a steep age--velocity dispersion relation \citep{dorman2015}, and a global burst of star formation in M31's disk 2\,--\,4 Gyr ago \citep{bernard2012, bernard2015, williams2017}, have motivated efforts to determine if a single major merger scenario, rather than a series of minor mergers, could produce M31's disturbed morphology.

Recent analysis of cosmological hydrodynamical simulations of M31 analogs \citep{hammer2018} suggests the possibility that a stream with the morphology of the GSS and a shell system like that observed in M31 could have been formed from a merger of M31 with a significantly more massive progenitor than implied from the numerical simulations of a minor merger. These simulations indicate that a major merger \citep[stellar mass ratio $< 4.5$ : 1,][]{hammer2018} occurring several Gyr ago could have formed not only the GSS and associated shells, but also built M31's inner halo, influenced the structure of the outer halo, and shaped the morphology of M31's disk. Moreover, a comparison of the halo properties of M31 with those of statistical ensembles of galaxies simulated in a cosmological volume indicates that such a major merger is statistically likely to have occurred \citep{dsouza2018}.

The measurements of \feh\ and \afe\ for stars in the GSS shed light on the environment in which these stars formed.
If the GSS's progenitor galaxy obeyed the mass--metallicity relation for dwarf galaxies in the Local Group \citep{kirby2013}, then it would have had a stellar mass of at least $\sim 0.5$\,--\,$2 \times 10^9~M_{\sun}$, which covers a range from slightly more massive than the present day stellar mass of M32 to slightly less massive than the LMC \citep{vandermarel2002,mcconnachie2012}.  However, this estimate is based on the median \feh\ of stars with successful \feh\ measurements in the GSS component (\hbox{$\sim-0.8$}, Figure~\ref{fig:mdfs}), which is a lower limit on the median \feh\ of the true distribution (Section~\ref{sec:mdf_biases}).

We can roughly estimate the magnitude of the bias introduced in the GSS MDF by the failed \feh\ measurements, as well as those removed based on the $\chi^2$ contours for $T_{\rm eff}$ and \feh, by assuming their \fehp\ values are accurate and computing the median (mean) \feh\ of stars with velocities within $2\sigma_v$ of the mean velocity of the GSS\@. This results in an estimate of $\langle {\rm [Fe/H]} \rangle = -0.68$ ($-0.74$) from 27 stars, 19 of which have spectroscopic \feh\ measurements.  The median \feh\ is likely more accurate, as the lower value for the mean is driven by one star with \feh$<-2.5$, and in general the stars in this field with the lowest \fehp\ estimates have significantly higher spectroscopic \feh\ estimates (Figures~\ref{fig:fehphot_vs_vel} and \ref{fig:data_feh_vs_vel}). An \feh\ of $\sim -0.68$ is consistent with a progenitor of higher stellar mass: $\sim 2^{+3}_{-1}\times 10^9M_{\sun}$ \citep[assuming an extension of the relation derived by][]{kirby2013},
fully consistent with the progenitor masses derived from the minor merger simulations.

While we cannot reliably estimate the effect of the exclusion of stars with TiO on the mean \feh\ of the GSS, a progenitor as massive as those invoked in the major merger scenarios \citep[e.g., roughly $M_*\sim 2.5\times 10^{10}$\Msun,][]{hammer2018,dsouza2018} would require $\langle {\rm [Fe/H]} \rangle \sim -0.05\pm 0.02$ to lie on the $z=0$ stellar mass--metallicity relation measured from SDSS galaxies (Leethochawalit et al., submitted). 
The eight stars with velocities consistent with the GSS velocity that were excluded due to the presence of TiO cannot drive the mean \feh\ of the GSS high enough to be even marginally consistent with this relation.  This remains true even for a smaller progenitor with a stellar mass of $10^{10}$\Msun.  

However, a star-forming progenitor of this size would be expected to have a metallicity gradient, and the SDSS metallicities from which the $z=0$ mass-metallicity relation are derived are measured in the central regions.  Thus, the \feh\ of the GSS is not {\it a priori} inconsistent with a major merger scenario if the stars in the GSS are not primarily from the center of the progenitor.  In \citeauthor{hammer2018}'s (\citeyear{hammer2018}) simulations of major mergers in M31-like halos, features similar to the GSS form from stars drawn from a large area within the progenitor, with 90\% of the stars drawn from a radial range of $\sim 5$\,--\,20~kpc (F.\ Hammer, private comm.). 
\citet{fardal2008} explored a minor merger scenario with a disk-galaxy progenitor with $M_*\sim 2\times 10^9$\Msun\ and a metallicity gradient consistent with that of M33.  In this simulation, the core of the GSS is dominated by relatively metal-rich stars that originated close to the center of the galaxy (within a couple disk scale lengths from the center; their Figures 1 and 2).  The metallicity of the GSS field presented here may therefore place useful observational constraints not only on the mass of the progenitor, but also on where in the progenitor the stream stars originated, regardless of whether the GSS was formed in a major or minor merger.  

\subsection{Origin of the KCC}
\defcitealias{gilbert2009gss}{G09}

\citet[hereafter referred to as G09]{gilbert2009gss} discussed in detail the potential origins of the second kinematically cold feature in this field, which is also seen in an adjoining spectroscopic field 
\citep[`H13s,' Figure~\ref{fig:roadmap};][]{kalirai2006gss}.  \citetalias{gilbert2009gss} ruled out an extended stellar disk origin for the feature, based in part on the constant ratio of stars in each component over the $\sim 30$~kpc in the disk plane covered by the two spectroscopic fields.  While an unrelated substructure could not be ruled out as the KCC's origin, \citetalias{gilbert2009gss} concluded that a direct physical association between the two features provided a natural explanation for the tight correlation of the mean velocity of the GSS and KCC as a function of position throughout the 7~kpc in projected distance from M31's center covered by the two fields, as well as the similarities in the CMD-distribution (\fehp) of the stars.

The similarity of the spectroscopic \feh\ distributions of the GSS and KCC as well as the consistency between the \afe\ abundances of the KCC and GSS components, provide further evidence for a potential physical link between the two features and simultaneously makes it more difficult to support an extended stellar disk origin for the KCC\@.  Possible GSS-related origins discussed by \citetalias{gilbert2009gss} include a bifurcation of the GSS and an extension of M31's western shelf.  While bifurcations in line-of-sight velocities can in principle be produced in tidal debris features, a possible impediment to this explanation is the large separation in velocity of the two observed features ($\sim 100$~\kms).  
None of the minor merger models of the encounter published to date have produced a bifurcation in the GSS that would lead to the observed signature.  

M31's western shelf is part of the forward continuation of the GSS\@.  Observations of M31's western shelf have measured the velocities of stars in this debris feature, which is a three dimensional shell \citep{fardal2012}.  The shell feature would need to extend nearly 180\degree\ in position angle in order to reach the position of these two GSS fields.  If it was perfectly symmetric, the velocities of stars in the KCC would be offset by $\sim -30$ to $-40$~\kms\ from the negative velocity caustic of the shell, and by $\sim -40$ to $-50$~\kms\ from the mean velocities of the stars aligned along the caustic \citep[Figure~8 of][]{fardal2012}.  Moreover, the stars observed in the western shelf primarily populate the positive velocity caustic.  There is no indication of a velocity peak corresponding to the positive velocity caustic (which would be expected at $\sim -200$~\kms\ in this GSS field). However, there are multiple potential deviations from symmetry that could be induced in the shell over a position angle this large.  

Recent simulations of major mergers in M31-like halos \citep{hammer2018} have introduced a third possibility: that the KCC is part of a previous wrap of the stream.  In this scenario, multiple loops of tidal debris (preceding and including the GSS) are viewed in projection along the loop plane.  \citet{hammer2018} proposed that multiple superposed loops of the stream could provide a natural explanation for the double peaked probability distributions found in some fields by \citet{conn2016}, who measured the line-of-sight distance of the GSS as a function of position along the stream.  Initial analysis of one of the \citet{hammer2018} models that includes multiple superposed loops indicates that a $\sim 100$~\kms\ offset in line of sight velocity between multiple loops can be produced (Y.\ Yang, private comm.).

While the \feh\ and \afe\ abundances of the KCC and GSS in this field provide additional support for a physical connection between the KCC and GSS, the origin of the KCC remains ambiguous.  There remains at least one viable GSS--related origin for the KCC in both the minor (extension of the western shelf) and major (a loop preceding the GSS) merger scenarios.  Additional modelling and analysis will be required to assess the viability of either of these potential origins.  However, if they remain viable, the assumption of either origin could be used in conjunction with the observed velocity and abundance distributions to place meaningful constraints on future models of the merger.

\section{Summary}\label{sec:conc}
We have presented the first \afe\ measurements, and the first \feh\ measurements derived from the strength of Fe lines via spectral synthesis, for stars in M31's GSS\@. We utilized previous kinematical modelling of this field to compare the \feh\ distributions of stars in each of the three M31 components present in the field: the GSS, a second component with a narrow velocity distribution (the KCC), and the underlying M31 stellar halo. The GSS and KCC have very similar \feh\ distributions, while the halo is slightly more metal-poor.  The similarity of the \feh\ and \afe\ distributions of the GSS and KCC provides additional support for the possibility that the KCC is physically related to the GSS\@.

The distribution of \afe\ as a function of \feh\ indicates that the stars in this field were drawn from an environment that experienced a higher efficiency of star formation than that of the surviving M31 dwarf satellites (at least those with abundance measurements): the stars in the GSS field on average have higher \feh\ and \afe\ abundances, and their \afe\ abundances decline with increasing \feh\ only at \feh\,$\sim -0.9$.  This indicates that the environment in which the stars in this field formed enriched to a higher metallicity than the comparison M31 satellite sample before Fe-rich Type~Ia supernovae began to explode.  Moreover, the steeply declining trend of \afe\ vs.\ \feh\ observed for \feh\,$\gtrsim -0.9$ implies that after enriching to this metallicity, the GSS's progenitor experienced a rapid decline in its SFR, allowing Type Ia supernovae to over-take core collapse supernovae as the dominant sites of nucleosynthesis.
Biases against the recovery of \feh\ and \afe\ for redder, and likely more metal-rich, stars  
prevent a definitive estimate of the stellar mass of the progenitor from the \feh\ measurements.  However, the abundances in the GSS field are consistent with a progenitor for the GSS that is at least as massive as that expected from simulations that reproduce the GSS and associated debris via a minor merger.

Initial comparisons with stars in other fields in M31's stellar halo indicate that stars in the GSS field, including the stars most likely associated with the underlying kinematically hot stellar halo component, are significantly more metal-rich than both a ``smooth'' inner halo field \citep[showing no evidence of substructure;][]{escala2019} and stars in M31's outer halo \citep{vargas2014apjl}.  The few stars with \afe\ and \feh\ measurements in M31's outer halo appear to be less $\alpha$-enhanced than stars in M31's inner halo.
However, the comparison sample sizes are currently too small to draw significant conclusions.  Future work will substantially increase the number of stars with abundance measurements in M31's GSS and associated debris, as well as in halo fields uncontaminated by substructure.

\acknowledgments
The authors recognize and acknowledge the very significant cultural role and reverence that the summit of Mauna Kea has always had within the indigenous Hawaiian community. We are most fortunate to have the opportunity to conduct observations from this mountain.  

The authors thank M.~Fardal, F.~Hammer, Y.~Yang, and S.~Hasselquist for helpful conversations during the writing of this manuscript, and A.~McConnachie for use of the PAndAS image.  This research made use of Astropy, a community-developed core Python package for Astronomy  \citep{astropy2013,astropy2018}\footnote{http://www.astropy.org}. 
  
Support for this work was provided by NSF grants AST-1614569 (K.M.G., J.W.), AST-1614081 (E.N.K., I.E.), and AST-1412648 (P.G.).  E.N.K. gratefully acknowledges support from a Cottrell Scholar award administered by the Research Corporation for Science Advancement as well as funding from generous donors to the California Institute of Technology.  I.E. acknowledges support from a National Science Foundation (NSF) Graduate Research Fellowship under Grant No.\ DGE-1745301. 
The analysis pipeline used to reduce the DEIMOS data was developed at UC Berkeley with support from NSF grant AST-0071048.

\facilities{Keck:II (DEIMOS), CFHT (MegaCam)}
\software{Astropy \citep{astropy2013},
Matplotlib \citep{matplotlib}, numpy \citep{numpy},
corner \citep{corner-sw_v1, corner}}


\bibliography{m31}

\appendix
\section{Velocity Model}\label{sec:app_velmodel}

The component-level analysis in Section~\ref{sec:mdfs} makes use of the Gaussian Mixture Model of stellar kinematics in M31's halo presented by \citet{gilbert2018}.  The velocity model includes all known M31 (halo and tidal debris features) and MW (thin and thick disk and halo) components observed in the SPLASH M31 halo dataset, and uses the M31 membership likelihoods discussed in Section~\ref{sec:data} (computed without the inclusion of the velocity diagnostic) as a prior on the probability that each star belongs to the MW or M31.  

Stellar kinematics, transformed to the Galactocentric frame, were fit in seven radial bins.  The velocity model for each bin includes both global parameters (describing the M31 halo, MW thin and thick disk, and MW halo) and field-specific parameters (describing the individual tidal debris features in each spectroscopic field), and assumes that each component is described by a Gaussian of mean velocity $\mu$ and velocity dispersion $\sigma$, and that each component contributes a fraction $f$ of stars to either the total M31 or MW populations.  A Markov-Chain Monte Carlo implementation \citep[{\tt emcee};][]{foreman-mackey_emcee, emcee-sw} was used to converge on the best-fit parameters, and marginalized posterior probability distribution functions for parameters of interest were derived from the converged MCMC chains.  

\begin{figure*}[tb!]
\includegraphics[width=\textwidth]{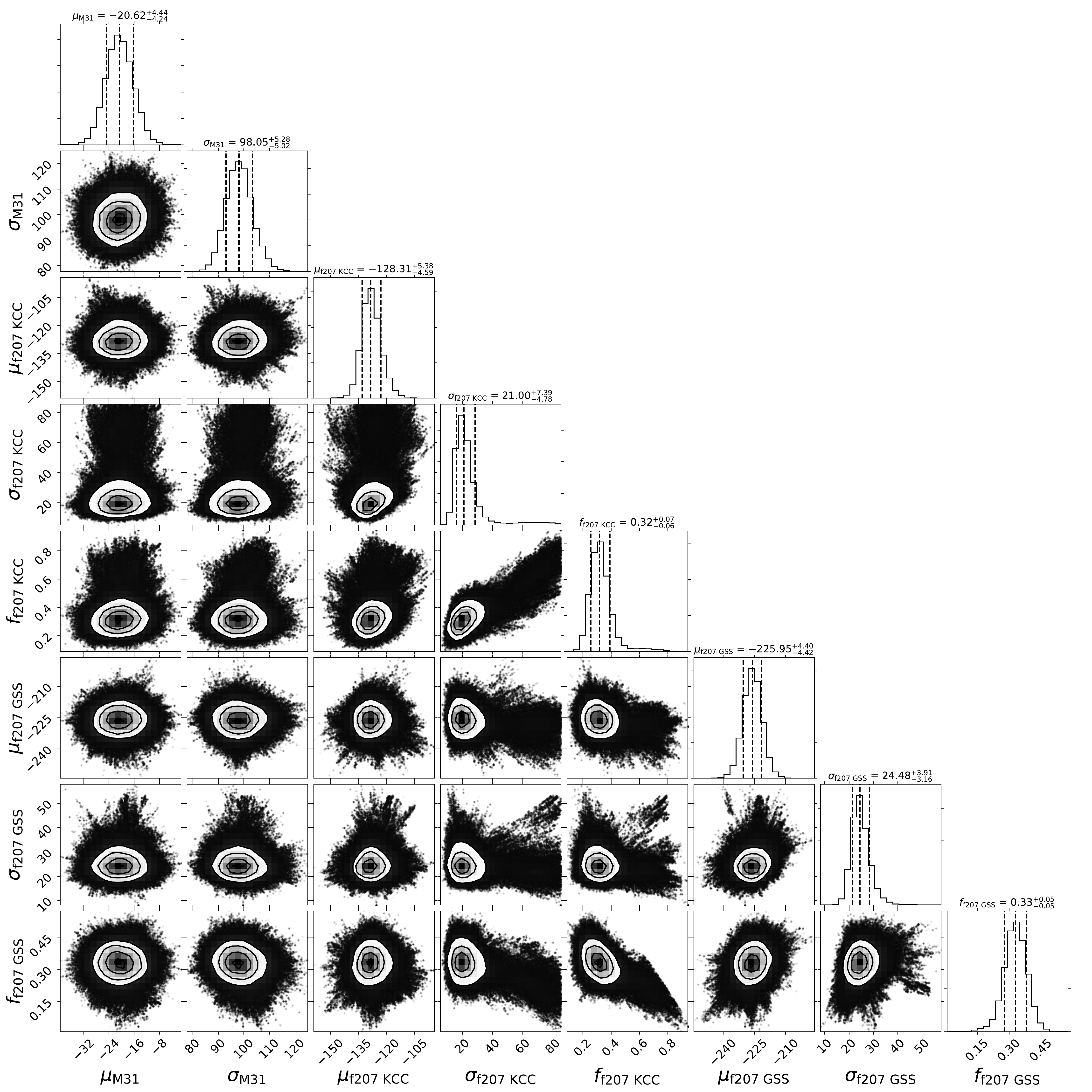}
\caption{
Marginalized one- and two-dimensional posterior probability distribution functions for each of the fit parameters relevant to this analysis \citep{gilbert2018}. Dashed lines and column headings show the 16th, 50th, and 84th percentiles of the marginalized 1-dimensional posterior probability distribution functions for each model parameter.  Velocities in this figure are in the Galactocentric frame.  While the parameters are well constrained by the model, several have low probability tails.  To capture the uncertainties inherent in assigning a velocity-based probability of membership in each component to a given star, we performed a series of random draws from the chains, iteratively deriving MDFs for each component for each random draw as described in Section~\ref{sec:mdfs}.  This figure was created using the open-source python package {\tt corner} \citep{corner,corner-sw_v2}.
}
\label{fig:corner}
\end{figure*}

The spectroscopic field analyzed here was included in the radial bin that encompassed stars located $14.1 \le$\rproj$< 24$~kpc in projection from M31's center.  
We drew from the \citet{gilbert2018} MCMC chains for that radial bin to explore the uncertainties in the component MDFs due to uncertainties in the velocity model.  Figure~\ref{fig:corner} shows the marginalized one-and two-dimensional posterior probability functions for all fit parameters used in Section~\ref{sec:mdfs}, including both global parameters and parameters specific to this spectroscopic field.  The velocities in Figure~\ref{fig:corner} are in the MW Galactocentric frame; line of sight mean velocities from the model are transformed back to the heliocentric frame (for the dataset presented here, $v_{\rm helio} = v_{\rm model} - 298.95$\kms) for the analysis described in Section~\ref{sec:compute_prob_mdfs}.  

\end{document}